\documentclass{article}
\usepackage{arxiv}

\usepackage{amsthm}

\usepackage{xcolor}  

\usepackage{amssymb}

\usepackage{amsmath}
\usepackage{mathdots}

\usepackage{mathtools}

\usepackage{tensor}

\usepackage{urwchancal}
\DeclareFontFamily{OT1}{pzc}{}
\DeclareFontShape{OT1}{pzc}{m}{it}{<-> s * [1.10] pzcmi7t}{}
\DeclareMathAlphabet{\mathpzc}{OT1}{pzc}{m}{it}

\usepackage{graphicx}
\usepackage{ifpdf}
\ifpdf
\usepackage{epstopdf}
\PrependGraphicsExtensions{.svg}
\fi

\newtheorem{theorem}{Theorem}[section]
\newtheorem{lemma}[theorem]{Lemma}
\newtheorem{corollary}[theorem]{Corollary}

\newtheorem{remark}[theorem]{Remark}

\providecommand{\R}{\mathbb{R}}

\providecommand{\SO}{\mathbf{SO}}

\providecommand{\GL}{\mathbf{GL}}
\providecommand{\SE}{\mathbf{SE}}

\providecommand{\SIM}{\mathbf{SIM}}

\providecommand{\grpG}{\mathbf{G}}

\providecommand{\Aut}{\mathbf{Aut}} 

\providecommand{\gothgl}{\mathfrak{gl}}

\providecommand{\gothg}{\mathfrak{g}}

\providecommand{\gothX}{\mathfrak{X}} 

\providecommand{\so}{\mathfrak{so}}
\providecommand{\se}{\mathfrak{se}}

\providecommand{\aut}{\mathfrak{aut}}

\providecommand{\Sph}{\mathrm{S}}

\providecommand{\calE}{\mathcal{E}}

\providecommand{\Sym}{\mathbb{S}} 

\providecommand{\tT}{\mathrm{T}} 

\providecommand{\eb}{\mathbf{e}} 

\DeclareMathOperator{\tr}{tr}

\DeclareMathOperator{\diag}{diag}

\DeclareMathOperator{\Ad}{Ad}

\providecommand{\Lyap}{\mathcal{L}} 

\providecommand{\td}{\mathrm{d}}
\providecommand{\tD}{\mathrm{D}}

\providecommand{\ddt}{\frac{\td}{\td t}}

\providecommand{\mr}[1]{\mathring{#1}} 

\usepackage{accents}
\usepackage{mathtools}
\makeatletter
\providecommand{\scirc}{%
    \hbox{\fontfamily{\rmdefault}\fontsize{0.4\dimexpr(\f@size pt)}{0}\selectfont{\raisebox{-0.52ex}[0ex][-0.52ex]{$\circ$}}}}

\makeatother

\makeatletter
\providecommand{\ucirc}{%
    \hbox{\fontfamily{\rmdefault}\fontsize{0.4\dimexpr(\f@size pt)}{0}\selectfont{\raisebox{0.0ex}[0ex][-0.52ex]{$\circ$}}}}

\makeatother

\mathchardef\mhyphen="2D

\providecommand{\etal}{\textit{et al.~}}

\usepackage{graphicx}
\usepackage{verbatim}
\usepackage{url}

\usepackage{tabularx}
\newcolumntype{x}[1]{>{\centering\arraybackslash\hspace{0pt}}p{#1}}
\usepackage{calc}
\usepackage{multirow}

\usepackage{pifont}
\newcommand{\tick}{{\color{green}$\checkmark$}}
\newcommand{\cross}{{\color{red} \ding{53}}}
\newcommand{\tsim}{{\color{blue}$\thicksim$}}

\newcommand{\set}[2]{\left\{ #1 \;\middle|\; #2 \right\}}
\renewcommand{\etal}{\textit{et al.~}}

\newcommand{\changeVersionOne}[1]{#1}
\newcommand{\changeVersionTwo}[1]{#1}

\usepackage{ifthen}

\newcommand{\makelongversion}{build the long version} 

\newenvironment{longversion}
{\ifthenelse{\isundefined{\makelongversion}}%
{\expandafter\comment}{}%
}{\ifthenelse{\isundefined{\makelongversion}}%
{\expandafter\endcomment}{}}
\newenvironment{shortversion}
{\ifdefined\makelongversion%
\expandafter\comment\fi%
}{\ifdefined\makelongversion%
\expandafter\endcomment\fi}

\usepackage{hyperref}

\newcommand{\papercitation}{
van Goor, Pieter, Tarek Hamel, and Robert Mahony. "Synchronous observer design for inertial navigation systems with almost-global convergence." Automatica 177 (2025): 112328.
}
\newcommand{\publicationdetails}
{
\papercitation \\
\copyrightNoticeIFACAccepted{2025} 
}
\newcommand{\publicationversion}
{Author Accepted Version}

\begin{document}




\headertitle{Synchronous Observer Design for Inertial Navigation Systems with Almost-Global Convergence}
\title{Synchronous Observer Design for Inertial Navigation Systems with Almost-Global Convergence}

\author{
\href{https://orcid.org/0000-0003-4391-7014}{\includegraphics[scale=0.06]{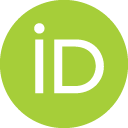}\hspace{1mm}
Pieter van Goor}
\\
    Robotics and Mechatronics group \\
    University of Twente \\
    7500 AE Enschede, The Netherlands \\
    \texttt{Pieter.vanGoor@anu.edu.au} \\
	\And	\href{https://orcid.org/0000-0002-7779-1264}{\includegraphics[scale=0.06]{orcid.png}\hspace{1mm}
    Tarek Hamel}
\\
    I3S (University C\^ote d'Azur, CNRS, Sophia Antipolis) \\
    and Insitut Universitaire de France \\
    \texttt{THamel@i3s.unice.fr} \\
	\And	\href{https://orcid.org/0000-0002-7803-2868}{\includegraphics[scale=0.06]{orcid.png}\hspace{1mm}
    Robert Mahony}
\\
    Systems Theory and Robotics Group \\
	Australian National University \\
    ACT, 2601, Australia \\
	\texttt{Robert.Mahony@anu.edu.au} \\
}

\maketitle

\vspace{1cm}

\begin{abstract}
An Inertial Navigation System (INS) is a system that integrates acceleration and angular velocity readings from an Inertial Measurement Unit (IMU), along with other sensors such as \changeVersionOne{Global Navigation Satellite Systems (GNSS)} position, GNSS velocity, and magnetometer, to estimate the attitude, velocity, and position of a vehicle.
This paper shows that the INS problem can be analysed using the automorphism group of the extended special Euclidean group $\SE_2(3)$: a group we term the \emph{extended similarity group} $\SIM_2(3)$.
By exploiting this novel geometric framework, we propose \changeVersionOne{a synchronous observer architecture; that is, an observer architecture for which the observer error is stationary if the correction terms are set to zero.}
In turn, this enables us to derive a modular, or plug-and-play, observer design for INS that allows different sensors to be added or removed depending on what is available in the vehicle sensor suite.
We prove both almost-global asymptotic and local exponential stability of the error dynamics for the common scenario of at least IMU and GNSS position.
To the authors' knowledge, this is the first non-linear observer design with almost global convergence guarantees or with plug-and-play modular capability.
A simulation with extreme initial error demonstrates the almost-global robustness of the system.
Real-world capability is demonstrated on data from a fixed-wing UAV, and the solution is compared to the state-of-the-art ArduPilot INS.
\end{abstract}

\section{Introduction}

Inertial Navigation Systems (INS) are algorithms that estimate a vehicle's \emph{navigation states} including its attitude, velocity, and position, with respect to a fixed reference frame.
The basic sensor used in INS is the Inertial Measurement Unit (IMU), consisting of a gyroscope and 3-axis accelerometer that measure the vehicle's angular velocity and specific acceleration, respectively.
If these measurements are exact and the initial states are exactly known, the dynamics of the vehicle can be forward-integrated to solve for the navigation states at any time.
In practice, noise from Micro Electrical Mechanical Systems (MEMS) hardware corrupts the measurements and leads such an estimation procedure to quickly diverge from the true states \cite{2007_woodman_IntroductionInertialNavigation}.
To prevent accumulation of error, real-world algorithms fuse information from additional sensors, typically \changeVersionOne{Global Navigation Satellite Systems (GNSS)} position, GNSS velocity, and magnetometer, into the state estimates.
The importance of reliable and accurate estimation of navigation states for aerospace, maritime, and robotics applications cannot be overstated, and continues to drive significant research into INS solutions.

\begin{table*}[!t]
    \changeVersionOne{
    \centering
    \caption{Comparison of qualitative properties of INS observer classes in the existing literature along with the proposed observer ``Ours''.
    The classes include Invariant Extended Kalman Filter (IEKF), Velocity Aided Attitude (VAA), and Nonlinear Integration Observer (NIO).
    Note that only stability domain characteristics are listed, as every observer in the table features local exponential stability.
    }
    \begin{tabularx}{\textwidth}{|X|x{\widthof{Constructive, Synchronous }}|x{\widthof{\tick (almost-global)}}|x{\widthof{\tsim (mag. required)}}|x{\widthof{GNSS}}|x{\widthof{GNSS}}|x{\widthof{GNSS}}|}
        \hline
        \multirow{2}{*}{Observer class} & \multirow{2}{*}{Methodology} & \multirow{2}{*}{Global Stability} & \multirow{2}{*}{Modular?} & \multicolumn{3}{c|}{Compatible Sensors} 
        \\ \cline{5-7}
        &&&&GNSS Pos. & GNSS Vel. & Mag.
        \\ \hline
        IEKF~\cite{2007_bonnabel_LeftinvariantExtendedKalman,2009_bonnabel_InvariantExtendedKalman,2011_barczyk_InvariantExtendedKalman,2017_barrau_InvariantExtendedKalman} & Linearisation, Symmetry & \cross (local) &
        \tick&\tick&\tick&\tick
        \\
        VAA~\cite{2010_hua_AttitudeEstimationAccelerated,2017_hua_RiccatiNonlinearObserver} & Riccati, High-Gain & \tsim (semi-global)  &
        \cross&\cross&\tick&\tick
        \\
        NIO~\cite{2013_grip_NonlinearObserverGNSSaided,2013_dukan_IntegrationFilterAPS,2017_hansen_NonlinearObserverDesign,2019_hansen_NonlinearObserverTightly} & Constructive, High-Gain &
        \tsim (semi-global)
        &  \tsim (mag. required) &\tick&\tick&\tick
        \\
        Berkane~\etal~\cite{2019_berkane_PositionVelocityAttitude,2021_berkane_NonlinearNavigationObserver} & Constructive, High-Gain &
        \tsim (semi-global)
        & \tsim (mag. required) &\tick&\cross&\tick
        \\
        Ours & Constructive, Synchronous &
        \tick (almost-global)
        & \tick &\tick&\tick&\tick
        \\
        \hline
    \end{tabularx}
    \label{tab:comparison_observers}
    }
\end{table*}

Since the advent of MEMS hardware for small remotely piloted aerial systems (RPAS), the INS problem has seen great interest from the nonlinear observer community.
Initial work focused on local guarantees of convergence and demonstrated that observers exploiting geometric structure of the system generally outperformed generic nonlinear designs.
Early work by Vik and Fossen \cite{2001_vik_NonlinearObserverGPS} examined the velocity-aided attitude (VAA) problem: the sub-problem of INS where only velocity and attitude are considered and position estimation is ignored.
They assumed an external measurement of attitude is available alongside the IMU and GNSS measurements, and derived a nonlinear observer with a locally exponentially stable attitude error and a globally exponentially stable velocity error.
Several authors examined the use of extended Kalman filter (EKF) and unscented Kalman filter (UKF) solutions \cite{2005_george_TightlyCoupledINS,2006_li_LowcostTightlyCoupled,2009_hirokawa_LowCostTightlyCoupled} and reported good performance in real-world situations, although no guarantees of stability were provided.
Early work on the invariant extended Kalman filter (IEKF) \cite{2007_bonnabel_LeftinvariantExtendedKalman} was applied by Martin and Sala\"un \cite{2008_martin_InvariantObserverEarthVelocityAided} and Bonnabel \etal \cite{2009_bonnabel_InvariantExtendedKalman} to propose VAA solutions.
Barczyk and Lynch \cite{2011_barczyk_InvariantExtendedKalman} likewise applied the IEKF methodology to the full GNSS-aided INS problem.
Finally, recent work by Barrau and Bonnabel \cite{2017_barrau_InvariantExtendedKalman} applied the IEKF using the novel extended Special Euclidean group $\SE_2(3)$ and showed that the INS dynamics are group-affine in this case.
These authors' works clearly demonstrated that the IEKF showed improved practical performance over the EKF alternative by utilising the available system geometry.

The success of the almost-globally asymptotically stable complementary filter for attitude estimation \cite{2008_mahony_NonlinearComplementaryFilters} inspired interest in the design of nonlinear observers with more-than-local stability guarantees.
Grip \etal \cite{2012_grip_ObserversInterconnectedNonlinear} proposed a method for combining nonlinear and linear observers such that the resulting combination is exponentially stable under appropriate assumptions on the chosen gains and initial system conditions.
This \changeVersionOne{\emph{Nonlinear Integration Observer} (NIO)} was applied by Grip \etal in \cite{2013_grip_NonlinearObserverGNSSaided} to develop a semi-globally exponentially stable observer for INS with magnetometer and GNSS position measurements.
A further extension by Dukan and Sorensen \cite{2013_dukan_IntegrationFilterAPS} applied this method to underwater vehicles and included consideration of IMU biases.
Recently, Hansen \etal have extended the approach from \cite{2013_grip_NonlinearObserverGNSSaided} to compensate for time-delayed GNSS measurements \cite{2017_hansen_NonlinearObserverDesign} and to develop an improved RTK-GNSS INS solution \cite{2019_hansen_NonlinearObserverTightly}, both of which were shown to yield substantial performance improvements in real-world experiments.
Hua \cite{2010_hua_AttitudeEstimationAccelerated} proposed a semi-globally stable observer for VAA based on the complementary filter \cite{2008_mahony_NonlinearComplementaryFilters}, and showed that it is almost-globally stable when the vehicle's acceleration is constant.
Hua \etal \cite{2017_hua_RiccatiNonlinearObserver} recently extended this with a time-varying Riccati gain to improve the local convergence properties of the filter.
Roberts and Tayebi \cite{2011_roberts_AttitudeEstimationAccelerating} likewise provided two observer designs for VAA with semi-global stability guarantees and utilising magnetometer measurements.
Recently, Wang and Tayebi \cite{2018_wang_GloballyExponentiallyStable} proposed a globally stable hybrid observer for INS aided by landmark measurements rather than GNSS, and Berkane and Tayebi \cite{2019_berkane_PositionVelocityAttitude} proposed a semi-globally exponentially stable nonlinear observer for INS with magnetometer and generic position measurements.
This was extended by Berkane \etal \cite{2021_berkane_NonlinearNavigationObserver} who showed that the INS state is uniformly observable if and only if the measurement of position is persistently exciting, and provided experimental validation of their proposed observer.
These approaches were summarised in a tutorial by Wang and Tayebi \cite{2020_wang_ObserversDesignInertial}.
\changeVersionOne{Table \ref{tab:comparison_observers} provides a qualitative comparison of observer designs in the existing literature along with the proposed observer.}
In summary, the various solutions proposed by the nonlinear observer community have provided a range of locally and semi-globally stable observers for VAA and INS, and have contributed to a principled understanding of the INS problem.

\changeVersionOne{
In this paper, we propose a novel nonlinear observer design for INS aided by GNSS position and, optionally, GNSS velocity and magnetometer measurements.
The proposed design relies on exploiting the group-affine INS dynamics modelled on $\SE_2(3)$ through a recently developed \emph{synchronous} observer architecture \cite{2021_vangoor_AutonomousErrorConstructive,2023_vangoor_ConstructiveEquivariantObserver}.
Synchrony of the observer architecture is a property of the observer dynamics when the correction term is set to zero; that is, a property of the pre-observer or internal model of the observer architecture.
An observer is said to be synchronous if there is a choice of error between system and observer state that is stationary for the purely internal model dynamics of the observer.
This highly unusual property results in a modular observer design where various correction terms associated with different sensors can be added together freely without introducing undesirable coupling effects.
Moreover, the time evolution of a Lyapunov function constructed from the synchronous error depends linearly on the observer correction terms; there are no autonomous error dynamics to compensate or dominate.
This allows almost global stability analysis, limited only by topological obstructions inherent in the error space, at least for persistently exciting state trajectories.


This paper extends the authors' prior work on INS \cite{2023_vangoor_ConstructiveEquivariantObserverb}.
The main contributions are:
\begin{enumerate}
    \item We introduce the extended similarity group $\SIM_2(3)$: a new Lie group for studying INS problems. We show that it forms the automorphism group for $\SE_2(3)$ and exploit this symmetry to propose the first synchronous observer architecture for INS.
    \item We propose an observer design with almost-globally asymptotically and locally exponentially stable error dynamics using GNSS position measurements and optionally using GNSS velocity and magnetometer measurements.
    \item We demonstrate real-world performance as compared to a state-of-the-art (open-source) multiplicative EKF implementation and provide simulations demonstrating the almost-global convergence properties of the proposed observer%
    \footnote{Code: \url{https://github.com/pvangoor/synchronous_INS}}.
\end{enumerate}
To the best of the authors' knowledge, and in contrast to the existing designs featuring local \cite{2001_vik_NonlinearObserverGPS,2005_george_TightlyCoupledINS,2006_li_LowcostTightlyCoupled,2009_hirokawa_LowCostTightlyCoupled,2008_martin_InvariantObserverEarthVelocityAided,2009_bonnabel_InvariantExtendedKalman,2011_barczyk_InvariantExtendedKalman,2017_barrau_InvariantExtendedKalman}
or semi-global \cite{2012_grip_ObserversInterconnectedNonlinear,2013_grip_NonlinearObserverGNSSaided,2013_dukan_IntegrationFilterAPS,2019_hansen_NonlinearObserverTightly,2010_hua_AttitudeEstimationAccelerated,2017_hua_RiccatiNonlinearObserver,2011_roberts_AttitudeEstimationAccelerating,2018_wang_GloballyExponentiallyStable,2019_berkane_PositionVelocityAttitude,2021_berkane_NonlinearNavigationObserver,2020_wang_ObserversDesignInertial} exponential stability,
the proposed observer is the first to exhibit almost-globally asymptotic and locally exponential stable error dynamics.
}

The paper is organised into nine sections, including the introduction.
Mathematical preliminaries are provided in Section 2.
The INS problem is formally posed in Section 3, and Section 4 provides an interpretation of the system dynamics on a Lie group through the introduction of the automorphism group $\SIM_2(3)$ of $\SE_2(3)$.
The proposed observer design is detailed in Sections 5 and 6, and is validated in Sections 7 and 8 through simulations and real-world experiments.
Conclusions and some directions for future work are given in Section 9.

\section{Preliminaries}

\subsection{Matrix Algebra}

For any square matrices $A,B \in \R^{n\times n}$, the matrix commutator is defined by
\begin{align*}
    [A,B] &= AB - BA.
\end{align*}
The set of $n\times n$ symmetric positive definite matrices is denoted $\Sym_+(n)$.
For two symmetric matrices $A,B$ we write $A > B$ to mean $A-B \in \Sym_+(n)$.

The Euclidean inner product and norm are defined for vectors and matrices by
\begin{align*}
    \langle v, w \rangle &= v^\top w, &
    \langle A, B \rangle &= \tr(A^\top B), \\
    \vert v \vert^2 &= v^\top v, &
    \vert A \vert^2 &= \tr(A^\top A),
\end{align*}
for all $v, w \in \R^n$ and $A, B \in \R^{n \times m}$, where $\tr : \R^{m\times m} \to \R$ is the trace operator.
The operator norm of a matrix $A \in \R^{n \times m}$ is defined by
\begin{align*}
    \Vert A \Vert = \inf\set{c \geq 0 }{\vert A v \vert \leq c \vert v \vert, \text{for all } v \in \R^m},
\end{align*}
and is equal to the square root of the largest eigenvalue of $A^\top A$.
We write $\diag(k_1,...,k_n)$ to mean the $\R^{n\times n}$ matrix with diagonal entries $k_1,...,k_n$ and all other entries zero.
For matrices $K \in \Sym_+(n)$ and $A \in \R^{n\times m}$, define the weighted norm
\begin{align*}
    \vert A \vert^2_K
    = \langle A, K A \rangle
    = \tr(A^\top K A).
\end{align*}

The 2-sphere is defined by $\Sph^2 := \set{y \in \R^3}{ \vert y \vert = 1}$.
The skew operator $\cdot^\times : \R^3 \to \R^{3\times 3}$ is defined by
\changeVersionOne{
\begin{align}\label{eq:skew_matrix_dfn}
    \Omega^\times = \begin{pmatrix}
        \Omega_1 \\ \Omega_2 \\ \Omega_3
    \end{pmatrix}^\times &= \begin{pmatrix}
        0 & -\Omega_3 & \Omega_2 \\
        \Omega_3 & 0 & -\Omega_1 \\
        -\Omega_2 & \Omega_1 & 0
    \end{pmatrix}.
\end{align}}
Note that $\Omega^\times v = \Omega \times v$ for all $\Omega, v \in \R^3$.
\changeVersionOne{The standard basis vectors of $\R^3$ are denoted by $\eb_1, \eb_2, \eb_3$.}
We call a set of time-varying vectors $\mu_1(t),...,\mu_n(t) \in \R^3$ \emph{directionally persistently exciting} if there exist $\delta, T > 0$ such that, for all $t \geq 0$,
\begin{align}\label{eq:persistently_exciting}
\int_t^{t+T} \sum_{i=1}^n \left(\mu_i^\times(\tau)^\top \mu_i^\times(\tau) \right) \td\tau > \delta I_3.
\end{align}
This is closely related to the condition explored in \cite{2012_trumpf_AnalysisNonLinearAttitude} for observability of attitude from directional measurements.

\changeVersionOne{
An equilibrium point of a nonlinear system is said to be \emph{almost-globally asymptotically stable} if it is locally stable and attractive from any initial condition outside of an exception set of measure zero.
For a detailed mathematical description of almost-global stability, we refer the reader to \cite{2004_angeli_AlmostGlobalNotion}.
}

\subsection{Lie Groups}

For a detailed introduction to matrix Lie groups, the authors recommend \cite{2015_hall_LieGroupsLie}.
\changeVersionOne{The Lie group of $n\times n$ invertible matrices is denoted $\GL(n)$ and its Lie algebra is denoted $\gothgl(n)$.}
Let $\grpG \leq \GL(n)$ denote a matrix Lie group.
The associated Lie algebra is denoted $\gothg \leq \gothgl(n)$ and may be identified with the tangent space of $\grpG$ at the identity.
The adjoint action $\Ad : \grpG \times \gothg \to \gothg$ is defined by
\begin{align*}
    \Ad_X(U) &= X U X^{-1},
\end{align*}
for all $X \in \grpG$ and all $U \in \gothg$.

An automorphism of a Lie group $\grpG$ is a diffeomorphism $\sigma : \grpG \to \grpG$ such that
\begin{align*}
    \sigma(XY) &= \sigma(X) \sigma(Y),
\end{align*}
for all $X,Y \in \grpG$.
In this case, we write $\sigma \in \Aut(\grpG)$.
The set of automorphisms of $\grpG$ is denoted $\Aut(\grpG)$ and is itself a Lie group, with Lie algebra $\aut(\grpG) \subset \gothX(\grpG)$, \changeVersionOne{where $\gothX(\grpG)$ denotes the smooth vector fields on $\grpG$}.
A vector field $\gamma \in \gothX(\grpG)$ lies in $\aut(\grpG)$ if it satisfies
\begin{align*}
    \gamma(XY) &= X \gamma(Y) + \gamma(X) Y.
\end{align*}

The special orthogonal group $\SO(3)$ and its Lie algebra $\so(3)$ are defined by
\begin{align*}
    \SO(3)
    &= \set{R \in \R^{3\times 3}}{ R^\top R = I_3, \; \det(R) = 1}, \\
    \so(3)
    &= \set{\Omega^\times \in \R^{3\times 3}}{ \Omega \in \R^3}.
\end{align*}
This is a matrix Lie group $\SO(3) \leq \GL(3)$.

The extended special Euclidean group $\SE_2(3)$ and its Lie algebra $\se_2(3)$ are defined by \cite{2014_barrau_InvariantParticleFiltering}
\begin{align*}
    \SE_2(3)
    &= \set{\begin{pmatrix}
        R & V \\0_{2\times 3} & I_2
    \end{pmatrix}
    }{R \in \SO(3), \; V \in \R^{3\times 2}}, \\
    \se_2(3)
    &= \set{\begin{pmatrix}
        \Omega^\times & W \\0_{2\times 3} & 0_{2\times 2}
    \end{pmatrix}
    }{\Omega \in \R^3, \; W \in \R^{3\times 2}}.
\end{align*}
The sub-matrix $V \in \R^{3 \times 2}$ can be thought of as encoding both the velocity $v \in \R^3$ and position $p \in \R^3$
\[
V = \begin{pmatrix} v & p \end{pmatrix} \in \R^{3 \times 2}
\]
of a moving frame encoded by $\SE_2(3)$.

\section{Problem Description}

Let $\{B\}$ denote the reference frame of an inertial measurement unit (IMU) that is moving with respect to an inertial reference frame $\{0\}$.
The attitude, velocity, and position of $\{B\}$ with respect to $\{0\}$ are written $R \in \SO(3)$, $v \in \R^3$, and $p \in \R^3$, respectively.
Then the dynamics of the system are given by
    \begin{align}
        \label{eq:system_dynamics_explicit}
    \dot{R} &= R \Omega^\times, &
    \dot{v} &= R a + g, &
    \dot{p} &= v,
\end{align}
where $\Omega \in \R^3$ is the measured angular velocity, $a \in \R^3$ is the measured specific acceleration, and $g \in \R^3$ is the known gravity vector in the inertial frame $\{0 \}$ (typically $g \approx (0,0,9.81)$~m/s$^2$).

Suppose the vehicle is equipped with a GNSS collocated with the IMU.
Then the modelled GNSS position $y_p \in \R^3$ and velocity $y_v \in \R^3$ are given by
\begin{subequations}
\label{eq:gnss_measurement}
\begin{align}
    y_p &= h_p(R,v,p) = p, \label{eq:gnss_measurement_pos} \\
    y_v &= h_v(R,v,p) = v \label{eq:gnss_measurement_vel},
\end{align}
\end{subequations}
respectively.
Suppose the IMU also includes a magnetometer.
Then the measured magnetic field $y_m \in \Sph^2$ is modelled by
\begin{align} \label{eq:magnetometer_measurement}
    y_m = h_m(R,v,p) = R^\top \mr{y}_m,
\end{align}
where $\mr{y}_m \in \Sph^2$ is the reference magnetic field direction.

The problem considered in this paper is the design of an observer for the attitude $R$, velocity $v$, and position $p$ of $\{B\}$ with respect to $\{0\}$, using the input measurements $(\Omega, a)$ and a combination of the magnetometer and GNSS output measurements.

\section{Lie Group Interpretation}

The system dynamics \eqref{eq:system_dynamics_explicit} may be written as dynamics on the extended Special Euclidean group $\SE_2(3)$.

\begin{lemma} \label{lem:system_dynamics_lg}
Define
\begin{align}
    X = \begin{pmatrix}
        R & V \\ 0_{2\times 3} & I_2
    \end{pmatrix} \in \SE_2(3),
    \label{eq:state_representation_lg}
\end{align}
with $V = (v \; p) \in \R^{3\times 2}$.
Then the system dynamics \eqref{eq:system_dynamics_explicit} may be written as
\begin{align}
    \dot{X} = X U + G X + N X - X N,
    \label{eq:system_dynamics_lg}
\end{align}
where
\begin{align*}
    U &:= \begin{pmatrix}
        \Omega^\times & W_U \\
        0_{2\times 3} & 0_{2\times 2}
    \end{pmatrix}, &
    W_U &= \begin{pmatrix}
        a & 0_{3\times 1}
    \end{pmatrix}, \\
    G &:= \begin{pmatrix}
        0_{3\times 3} & W_G \\
        0_{2\times 3} & 0_{2\times 2}
    \end{pmatrix}, &
    W_G &= \begin{pmatrix}
        g & 0_{3\times 1}
    \end{pmatrix}, \\
    N &:= \begin{pmatrix}
        0_{3\times 3} & 0_{3\times 2} \\
        0_{2\times 3} & S_N
        \end{pmatrix}, &
    S_N &:= \begin{pmatrix}
        0 & -1 \\ 0 & 0
    \end{pmatrix}.
\end{align*}
\end{lemma}

\begin{longversion}
\begin{proof}
Expanding \eqref{eq:system_dynamics_lg} into components yields
\begin{align*}
    \dot{X}
    &= \begin{pmatrix}
        R \Omega^\times & R W_U + W_G - V S_N \\ 0_{2\times 3} & 0_{2\times 2}
    \end{pmatrix}.
\end{align*}
This shows that the proposed dynamics \eqref{eq:system_dynamics_explicit} lie in the tangent space of $\SE_2(3)$ at $X$, and, indeed, $\dot{R} = R \Omega^\times$.
For the position and velocity terms, one has that
\begin{align*}
    \ddt \begin{pmatrix} v & p \end{pmatrix}
    &=R \begin{pmatrix} a & 0_{3\times 1} \end{pmatrix}
    + \begin{pmatrix} g \; 0_{3\times 1} \end{pmatrix}
    - \begin{pmatrix} v & p \end{pmatrix}
    \begin{pmatrix} 0 & -1 \\ 0 & 0 \end{pmatrix}, \\
    &=\begin{pmatrix} R a & 0_{3\times 1} \end{pmatrix}
    + \begin{pmatrix} g \; 0_{3\times 1} \end{pmatrix}
    + \begin{pmatrix} 0 & v \end{pmatrix}, \\
    &=\begin{pmatrix} R a + g && v \end{pmatrix},
\end{align*}
which matches the explicit system dynamics \eqref{eq:system_dynamics_explicit}.
\end{proof}
\end{longversion}

\begin{shortversion}
\changeVersionOne{
The result is obtained by expanding \eqref{eq:system_dynamics_lg} and matching the component terms of the left- and right-hand sides.
A detailed proof is provided in the arxiv version \cite{2023_vangoor_SynchronousObserverDesign}.
}
\end{shortversion}

The matrix differential equation \eqref{eq:system_dynamics_lg} does not fit the standard form of left-, right-, or mixed-invariant systems, as $N \not\in \se_2(3)$ \cite{2007_bonnabel_LeftinvariantExtendedKalman,2016_khosravian_StateEstimationInvariant}.
Interestingly, however, these equations can still be exactly integrated for constant input $U$ \cite{2019_eckenhoff_ClosedformPreintegrationMethods,2019_henawy_AccurateIMUPreintegration}.
This property is associated with a more fundamental structure where the matrix $N$ is associated with a time varying automorphism of the group element $X \in \SE_2(3)$, and indeed the integration admits a closed form using the matrix exponential \cite{2021_vangoor_AutonomousErrorConstructive}.

\subsection{Automorphisms of $\SE_2(3)$}

For a deep discussion of Lie group automorphisms, the authors recommend \cite{1965_hochschild_StructureLieGroups}.
We define the extended similarity group $\SIM_2(3)$ and its Lie algebra $\mathfrak{sim}_2(3)$ by
\begin{align*}
    \SIM_2(3) &=
    \set{
    \begin{pmatrix}
        R & V \\ 0_{2\times 3} & A
    \end{pmatrix}
    }{
    R \in \SO(3),
    V \in \R^{3\times 2},
    A \in \GL(2)
    }, \\
    \mathfrak{sim}_2(3) &=
    \set{
    \begin{pmatrix}
        \Omega^\times & W \\ 0_{2\times 3} & S
    \end{pmatrix}
    }{
    \Omega \in \R^3, \;
    W \in \R^{3\times 2}, \;
    S \in \gothgl(2)
    }.
\end{align*}
It is easily verified that this is a matrix Lie group $\SIM_2(3) \leq \GL(5)$.
The extended special Euclidean group $\SE_2(3)$ is a subgroup associated with the restriction $A = I_2$ and the extended special Euclidean algebra is a subalgebra associated with the restriction $S = 0_{2 \times 2}$.

\begin{lemma}\label{lem:conjugation_by_sim23}
The matrix group $\SE_2(3)$ is closed under conjugation by $\SIM_2(3)$.
That is, for any $Z \in \SIM_2(3)$ and $X \in \SE_2(3)$, then $Z^{-1} X Z \in \SE_2(3)$.
\end{lemma}

\begin{longversion}
\begin{proof}
Let $Z \in \SIM_2(3)$ and $X \in \SE_2(3)$ be arbitrary, and write
\begin{align*}
    Z &= \begin{pmatrix}
        R_Z & V_Z \\ 0_{2\times 3} & A_Z
    \end{pmatrix}, &
    X &= \begin{pmatrix}
        R_X & V_X \\ 0_{2\times 3} & I_2
    \end{pmatrix}.
\end{align*}
Then the conjugate may be computed as
\begin{align}
    Z^{-1}XZ
    &=\begin{pmatrix}
        R_Z & V_Z \\ 0_{2\times 3} & A_Z
    \end{pmatrix}^{-1}
    \begin{pmatrix}
        R_X & V_X \\ 0_{2\times 3} & I_2
    \end{pmatrix}
    \begin{pmatrix}
        R_Z & V_Z \\ 0_{2\times 3} & A_Z
    \end{pmatrix}, \notag\\
    &= \begin{pmatrix}
        R_Z^\top & - R_Z^\top V_Z A_Z^{-1} \\
        0_{2\times 3} & A_Z^{-1}
    \end{pmatrix}
    \begin{pmatrix}
        R_X R_Z & V_X A_Z + R_X V_Z \\
        0_{2\times 3} & A_Z
    \end{pmatrix}, \notag\\
    &= \begin{pmatrix}
        R_Z^\top R_X R_Z &&
        R_Z^\top (V_X A_Z + R_X V_Z) - R_Z^\top V_Z A_Z^{-1} A_Z \\
        0_{2\times 3} &&
        A_Z^{-1} A_Z
    \end{pmatrix}, \notag\\
    &= \begin{pmatrix}
        R_Z^\top R_X R_Z &&
        R_Z^\top (V_X A_Z + R_X V_Z - V_Z) \\
        0_{2\times 3} &&
        I_2
    \end{pmatrix}.
    \label{eq:conjugation_expanded}
\end{align}
Since $R_Z^\top R_X R_Z \in \SO(3)$ clearly, it follows that $Z^{-1}XZ \in \SE_2(3)$, as required.
\end{proof}
\end{longversion}

\begin{shortversion}
A detailed proof, based on expanding the matrix expressions, is provided in the arxiv version \cite{2023_vangoor_SynchronousObserverDesign}.
\end{shortversion}
Lemma~\ref{lem:conjugation_by_sim23} shows that each $Z \in \SIM_2(3)$ induces an automorphism of $\SE_2(3)$ by conjugation.
This is the key insight that facilitates the observer design undertaken in the sequel.
The following corollary \begin{shortversion}\cite{2023_vangoor_SynchronousObserverDesign}\end{shortversion}
follows by differentiating the conjugation $Z^{-1} X Z$ at the identity $Z = I_5$.

\begin{corollary}
    For all $N \in \mathfrak{sim}_2(3)$ and $X \in \SE_2(3)$, the matrix $N X - X N$ is an element of $\tT_X \SE_2(3)$.
\end{corollary}

\section{Observer Design}
\label{sec:observer_design}

\subsection{Observer Architecture}

The generic observer architecture for group affine systems proposed in \cite{2021_vangoor_AutonomousErrorConstructive} is adapted here to the special case of the IMU system \eqref{eq:system_dynamics_explicit}.
The interpretation of \eqref{eq:system_dynamics_explicit} as dynamics on a Lie group in Lemma \ref{lem:system_dynamics_lg}, coupled with the relationship between $\SIM_2(3)$ and the automorphisms of $\SE_2(3)$ leads to the following synchronous observer architecture.

\begin{lemma} \label{lem:observer_architecture}
Let $X \in \SE_2(3)$ be defined as in Lemma \ref{lem:system_dynamics_lg}.
Define the state estimate $\hat{X} \in \SE_2(3)$ and auxiliary observer state $Z \in \SIM_2(3)$ to have dynamics
\begin{subequations}\label{eq:observer_architecture}
\begin{align}
    \dot{\hat{X}} &= \hat{X} U + G \hat{X} + N \hat{X} - \hat{X} N + \Ad_Z(\Delta) \hat{X}, \label{eq:dothatX}\\
    \dot{Z} &= (G + N) Z - Z \Gamma, \label{eq:dotZ}
\end{align}
\end{subequations}
where $\Delta \in \se_2(3)$ and $\Gamma \in \mathfrak{sim}_2(3)$ are correction terms that remain to be designed.
Define the observer error
\begin{align} \label{eq:error_definition}
    \bar{E} := Z^{-1} X \hat{X}^{-1} Z \in \SE_2(3).
\end{align}
\changeVersionOne{Then the observer and system are $\bar{E}$-synchronous \cite{2010_lageman_GradientLikeObserversInvariant}: the dynamics of the error term $\bar{E}$ depend only on the chosen correction terms $\Delta, \Gamma$, and, in particular,}
\begin{align} \label{eq:error_dynamics}
    \dot{\bar{E}} = \Gamma \bar{E} - \bar{E} \Gamma - \bar{E} \Delta.
\end{align}
\end{lemma}

\begin{longversion}
\begin{proof}
It is straightforward to verify that the observer architecture respects the Lie group structures of $\SE_2(3)$ and $\SIM_2(3)$, and hence $\hat{X} \in \SE_2(3)$ and $Z \in \SIM_2(3)$ for all time.
The error dynamics \eqref{eq:error_definition} are obtained by computing
\begin{align*}
    \dot{\bar{E}}
    &= \ddt Z^{-1} X \hat{X}^{-1} Z, \\
    &= -(Z^{-1} (G + N) - \Gamma Z^{-1} ) X \hat{X}^{-1} Z
    \\ &\hspace{1cm}
    + Z^{-1} (XU + GX + NX - XN) \hat{X}^{-1} Z
    \\ &\hspace{1cm}
    - Z^{-1} X (U \hat{X}^{-1} + \hat{X}^{-1} G + \hat{X}^{-1} N - N \hat{X}^{-1}
    + \hat{X}^{-1} \Ad_Z (\Delta)) Z
    \\ &\hspace{1cm}
    + Z^{-1} X \hat{X}^{-1} ((G+N)Z - Z \Gamma)
    , \\
    &= \Gamma Z^{-1} X \hat{X}^{-1} Z
    + Z^{-1} X \hat{X}^{-1} \Ad_Z (\Delta) Z
    - Z^{-1} X \hat{X}^{-1} Z \Gamma
    , \\
    &= \Gamma Z^{-1} X \hat{X}^{-1} Z - Z^{-1} X \hat{X}^{-1} Z \Gamma
    + Z^{-1} X \hat{X}^{-1} Z \Delta
    , \\
    &=  \Gamma \bar{E} - \bar{E} \Gamma - \bar{E} \Delta.
\end{align*}
The synchrony of the observer and system is verified by noting that $\dot{\bar{E}} = 0$ when $\Delta = 0$ and $\Gamma = 0$.
The synchronicity of the observer and system is verified by noting that $\dot{\bar{E}} = 0$ when $\Delta = 0$ and $\Gamma = 0$.
\end{proof}
\end{longversion}

\begin{shortversion}
The proof is obtained by direct differentiation of the error dynamics and is provided in the arxiv version \cite{2023_vangoor_SynchronousObserverDesign}.
\changeVersionOne{Lemma \ref{lem:system_dynamics_lg} is used to expand the dynamics of $X$ into the matrix form \eqref{eq:system_dynamics_lg}, and Lemma \ref{lem:conjugation_by_sim23} is necessary to ensure that the error $\bar{E}$ lies in $\SE_2(3)$.}
\end{shortversion}

\changeVersionOne{
Synchrony is a powerful property of an observer architecture that facilitates global analysis of error dynamics.
Since the error dynamics depend linearly on the correction terms, the error can be kept stationary by choosing the correction terms to be zero.
In the more common asynchronous observer design problems, the error dynamics feature exogenous dynamics relating to the mismatch between the observer and system state, which must be overcome to ensure convergence and stability of the error.
This is often done through linearisation of the error dynamics, but this restricts the validity of the resulting observer design to a local domain.
In the proposed design, the auxiliary state $Z \in \SIM_2(3)$ acts to cancel the exogenous dynamics of $E = X \hat{X}^{-1}$, which is the asynchronous error term that is more commonly studied in related prior work \cite{2016_khosravian_StateEstimationInvariant,2019_henawy_AccurateIMUPreintegration,2020_wang_ObserversDesignInertial}.
}

The following two lemmas are obtained by expanding the matrix expressions for the auxiliary state matrix $Z$ and the error matrix $\bar{E}$.
\changeVersionOne{They are used throughout Section \ref{sec:observer_design} to determine the effect of various correction terms on the components of $Z$ and $\bar{E}$.}
\begin{shortversion}
Full derivations are provided in the arxiv version \cite{2023_vangoor_SynchronousObserverDesign}.
\end{shortversion}

\begin{lemma}\label{lem:auxiliary_state_dynamics}
Consider the auxiliary state $Z$ with dynamics \eqref{eq:dotZ} as specified in Lemma \ref{lem:observer_architecture}.
Let $R_Z \in \SO(3), V_Z \in \R^{3\times 2}, A_Z \in \GL(2)$ denote the rotation, translation, and scaling components of $Z$, respectively.
Then their dynamics are given by
\begin{subequations}
\begin{align}
    \dot{R}_Z &= -R_{Z} \Omega_\Gamma^\times,
    \label{eq:expanded_auxiliary_dynamics_R} \\
    \dot{V}_Z &= W_G A_Z - R_Z W_\Gamma - V_Z S_\Gamma,
    \label{eq:expanded_auxiliary_dynamics_V} \\
    \dot{A}_Z &= S_N A_Z - A_Z S_\Gamma.
    \label{eq:expanded_auxiliary_dynamics_A}
\end{align}
\end{subequations}
\end{lemma}

\begin{lemma}\label{lem:error_state_and_dynamics}
Consider the observer error $\bar{E}$ as defined under the conditions of Lemma \ref{lem:observer_architecture}.
Let $R_{\bar{E}}\in \SO(3)$ and $ V_{\bar{E}} \in \R^{3\times 2}$ denote its rotation and translation components, respectively.
Then,
\begin{subequations}
    \begin{align}
        R_{\bar{E}} &= R_Z^\top R \hat{R}^\top R_Z,
        \label{eq:expanded_error_R} \\
        V_{\bar{E}}
        &= R_Z^\top(V A_Z - V_Z) - R_{\bar{E}} R_Z^\top (\hat{V} A_Z - V_Z).
        \label{eq:expanded_error_V}
    \end{align}
\end{subequations}
The dynamics of these components are given by
\begin{subequations}
    \begin{align}
        \dot{R}_{\bar{E}} &= \Omega_\Gamma^\times R_{\bar{E}} - R_{\bar{E}}(\Omega_\Gamma^\times + \Omega_\Delta^\times),
        \label{eq:expanded_error_dynamics_R} \\
        \dot{V}_{\bar{E}} &= \Omega_\Gamma^\times V_{\bar{E}} - V_{\bar{E}} S_\Gamma + (I - R_{\bar{E}})W_\Gamma - R_{\bar{E}} W_\Delta.
        \label{eq:expanded_error_dynamics_V}
    \end{align}
\end{subequations}
\end{lemma}

The goal in our observer design is to drive $\bar{E} \to I_5$ (see \cite{2021_vangoor_AutonomousErrorConstructive}), noting that
$\bar{E} = I_5$ if and only if $X = \hat{X}$.
We approach the design problem using the Lyapunov observer design methodology.
Let $\Lyap : \SE_2(3) \to \R^+$ be a candidate cost function.
It follows from Lemma \ref{lem:observer_architecture} that
\begin{align} \label{eq:lyapunov_derivative}
    \dot{\Lyap} = \tD_{\bar{E}} \Lyap(\bar{E})[\Gamma \bar{E} - \bar{E} \Gamma - \bar{E} \Delta].
\end{align}
Then, for a given collection of sensors, we wish to find correction terms $(\Delta, \Gamma)$ for which the right hand side of \eqref{eq:lyapunov_derivative} is less or equal to zero.
This is an algebraic problem and is typically solved by exploiting the full sensor suite in a single, highly coupled design process.
The resulting correction terms can only be applied if all the sensors are available.
In contrast to this approach, we show that we can design separate ``modular'' correction terms for each sensor, each of which ensures non-increase of $\Lyap$, but none of which by themselves ensure convergence of $\Lyap$.
\changeVersionTwo{We show that by designing each of these terms correctly and exploiting synchrony, they may be added together to ensure the decrease of $\Lyap$, and lead to almost-global asymptotic and local exponential stability of the system.
As a result, any sensor can be plugged in or taken out at any time without compromising the stability or robustness of the observer design, although a minimal sensor suite is required to ensure observability and achieve full convergence of $\Lyap$ (see Section \ref{sec:full_observer_designs}).}


\begin{theorem}\label{thm:combined_corrections}
Let $\Lyap : \SE_2(3) \to \R^+$ be a uniformly continuous positive-definite cost function \changeVersionOne{with minimum $\Lyap(I_5) = 0$}.
Let $(\Delta_i, \Gamma_i) \in \se_2(3) \times \mathfrak{sim}_2(3)$ be a
collection of correction terms and define the component total-derivatives of $\Lyap$ by
\begin{align*}
    \dot{\Lyap}^i := \tD_{\bar{E}} \Lyap(\bar{E})[\Gamma_i \bar{E} - \bar{E} \Gamma_i - \bar{E} \Delta_i],
\end{align*}
for each $i=1,\dots,n$.
Suppose that $\dot{\Lyap}^i \leq 0$ and that $\Delta_i$ and $\Gamma_i$ are bounded and uniformly continuous in time.
Define the correction terms
\begin{align*}
    \Delta &= \sum_{i=1}^n \alpha_i \Delta_i, &
    \Gamma &= \sum_{i=1}^n \alpha_i \Gamma_i,
\end{align*}
for some (possibly time-varying) uniformly continuous gains $\alpha_i \geq \underline{\alpha} > 0$.
Then the cost function $\Lyap$ is a Lyapunov function for the dynamics of $\bar{E}$, in the sense that $\dot{\Lyap} \leq 0$, and its set of equilibria is exactly the intersection $\calE := \cap_{i=1}^n \calE_i$, where
\begin{align*}
    \calE_i :=
    \set{\bar{E} \in \SE_2(3)}{\dot{\Lyap}^i \equiv 0}.
\end{align*}
\changeVersionOne{Moreover, $\calE$ is nonempty and contains at least the global minimum $I_5$ of $\Lyap$. }
\end{theorem}

\begin{proof}
By $\bar{E}$-synchrony and linearity of the derivative, it follows that
\begin{align*}
    \dot{\Lyap}(\bar{E})
    &= \tD_{\bar{E}} \Lyap(\bar{E})[\Gamma \bar{E} - \bar{E} \Gamma - \bar{E} \Delta] 
    = \sum_{i=1}^n \alpha_i \dot{\Lyap}^i
    \leq \underline{\alpha} \sum_{i=1}^n \dot{\Lyap}^i.
\end{align*}
It follows immediately that $\dot{\Lyap} \leq 0$, and thus $\Lyap$ is indeed a Lyapunov function for $\bar{E}$.
Barbalat's lemma may be applied to obtain the equilibria of $\Lyap$.
Since $\dot{\Lyap} \leq 0$, it follows that $\Lyap \to \Lyap_{\lim} < \infty$.
Then $\dot{\Lyap} \to 0$ since $\dot{\Lyap}$ is the composition of $\tD \Lyap$ with the sum of $\Delta_i$ and $\Gamma_i$ terms, which are uniformly continuous in time.
The equilibria of $\Lyap$ are therefore exactly those points $\bar{E}$ for which $\sum_{i=1}^n \alpha_i \dot{\Lyap}^i \equiv 0$.
Since each individual summand is non-positive, it follows that $\dot{\Lyap} \equiv 0$ only when $\dot{\Lyap}^i \equiv 0$ for each $i = 1,..., n$.
This completes the proof.
\end{proof}

Theorem \ref{thm:combined_corrections} exhibits the modularity or plug-and-play nature of the correction terms.
Any correction terms for which the cost $\Lyap$ is decreasing may be added together, and doing so will only improve the performance.
This is particularly useful when considering correction terms that use different sensors.
For instance, suppose one has designed a correction term for a GNSS position measurement and separately designed a correction term for a magnetometer measurement.
According to the theorem above, these correction terms may simply be added together in a combined correction term without introducing negative impacts by their coupling.
In practice, however, noise in various sensor measurements will require careful tuning of gains to control the rate and robustness of the observer convergence.
In Section \ref{sec:full_observer_designs}, we apply the theorem to combine several correction terms built from different sensors, including GNSS position, GNSS velocity, and magnetometer measurements.

\subsection{Design of Individual Correction Terms}

The observer architecture of Lemma \ref{lem:observer_architecture} leaves the correction terms $\Gamma$ and $\Delta$ to be designed from the measurements \eqref{eq:gnss_measurement} and \eqref{eq:magnetometer_measurement}.
The strategy followed in this paper is to define several correction terms that independently cause a candidate Lyapunov function to decrease.
Theorem \ref{thm:combined_corrections} then ensures that a sum of these correction terms stabilises the Lyapunov function.

Let $R_{\bar{E}} \in \SO(3)$ and $V_{\bar{E}} \in \R^{3\times 2}$ denote the rotational and translational components of the error $\bar{E} \in \SE_2(3)$ defined in \eqref{eq:error_definition}.
Define the cost function
\begin{align} \label{eq:lyapunov_function}
    \Lyap(\bar{E}) := \tr(I_3 - R_{\bar{E}}) + \vert V_{\bar{E}} \vert^2.
\end{align}
Let $\Omega_\Delta, W_\Delta, \Omega_\Gamma, W_\Gamma, S_\Gamma$ denote the individual components of the correction terms $\Delta \in \se_2(3)$ and $\Gamma \in \mathfrak{sim}_2(3)$; that is,
\begin{align*}
    \Delta &= \begin{pmatrix}
        \Omega_\Delta^\times & W_\Delta \\ 0_{2\times 3} & 0_{2\times 2}
    \end{pmatrix}, &
    \Gamma &= \begin{pmatrix}
        \Omega_\Gamma^\times & W_\Gamma \\ 0_{2\times 3} & S_\Gamma
    \end{pmatrix}.
\end{align*}
Then
\begin{align}
    &\dot{\Lyap}(\Delta, \Gamma)
    = \tD_{\bar{E}} \Lyap(\bar{E})[\Gamma \bar{E} - \bar{E} \Gamma - \bar{E} \Delta], \label{eq:correction_derivative_expansion}\\
    &= \tr(R_{\bar{E}} \Omega_\Delta^\times)
    + 2\langle V_{\bar{E}}, - V_{\bar{E}} S_\Gamma + (I - R_{\bar{E}})W_\Gamma - R_{\bar{E}} W_\Delta \rangle.\notag
\end{align}

\subsubsection{Correction Terms using GNSS Position}

The following lemmas develop correction terms $\Delta$ and $\Gamma$ that rely on the measurement of position provided by a GNSS.
Lemma \ref{lem:gps_correction_V} gives correction terms that affect only the velocity and position error, and Lemma \ref{lem:gps_correction_R} gives correction terms that also affect the attitude error.

\begin{lemma}\label{lem:gps_correction_V}
Consider the cost function $\Lyap$ as defined in \eqref{eq:lyapunov_function}.
Choose gains $k_p > 0, K_q \in \Sym_+(2)$ and define the correction terms
\begin{align*}
    \Omega_\Delta &= 0  , \\
    W_\Delta &= k_p R_Z^\top (y - \hat{y}) C_p^\top A_Z^{-\top}, \\
    \Omega_\Gamma &= 0, \\
    W_\Gamma &= -k_p  R_Z^\top (y - V_Z A_Z^{-1} C_p)C_p^\top A_Z^{-\top}, \\
    S_\Gamma &= -\frac{k_p}{2} A_Z^{-1} C_p C_p^\top A_Z^{-\top} + \frac{1}{2} A_Z^\top K_q A_Z,
\end{align*}
where $y = h_p(X)$ and $\hat{y} = h_p(\hat{X})$ are the true and estimated position measurements \eqref{eq:gnss_measurement_pos}, and $C_p^\top = (0 \; 1) \in \R^{1 \times 2}$.
Then
\begin{align}
    \dot{\Lyap}(\Delta, \Gamma) &= - k_p \vert V_{\bar{E}} A_Z^{-1} C_p \vert^2 - \vert A_Z V_{\bar{E}}^\top \vert_{K_q}^2. \label{eq:lyapunov_derivative_gnss_pos1}
\end{align}
\end{lemma}

\begin{proof}
Substituting the chosen correction terms $W_\Gamma$ and $W_\Delta$ into the relevant terms of \eqref{eq:correction_derivative_expansion} yields
\begin{align*}
    (I - R_{\bar{E}})W_\Gamma - R_{\bar{E}} W_\Delta
    &= -k_p (I - R_{\bar{E}}) R_Z^\top (y - V_Z A_Z^{-1} C_p)C_p^\top A_Z^{-\top}
    \\ &\hspace{1cm}
    - k_p R_{\bar{E}} R_Z^\top (y - \hat{y}) C_p^\top A_Z^{-\top}, \\
    &= -k_p R_Z^\top (y - V_Z A_Z^{-1} C_p)C_p^\top A_Z^{-\top}
    \\ &\hspace{1cm}
    + k_p R_{\bar{E}} R_Z^\top (\hat{y} - V_Z A_Z^{-1} C_p)C_p^\top A_Z^{-\top}, \\
    &= -k_p R_Z^\top (V C_p - V_Z A_Z^{-1} C_p)C_p^\top A_Z^{-\top}
    \\ &\hspace{1cm}
    + k_p R_{\bar{E}} R_Z^\top (\hat{V} C_p  - V_Z A_Z^{-1} C_p)C_p^\top A_Z^{-\top}, \\
    &= - k_p R_Z^\top (V A_Z - V_Z)A_Z^{-1} C_p C_p^\top A_Z^{-\top}
    \\ &\hspace{1cm}
    + k_p R_{\bar{E}} R_Z^\top (\hat{V} A_Z  - V_Z) A_Z^{-1} C_p C_p^\top A_Z^{-\top}, \\
    &= -k_p V_{\bar{E}} A_Z^{-1} C_p C_p^\top A_Z^{-\top}.
\end{align*}
Using this and substituting the remaining correction terms into \eqref{eq:correction_derivative_expansion} yields
\begin{align*}
    \dot{\Lyap}(\Delta, \Gamma)
    &= - 2\langle V_{\bar{E}}^\top V_{\bar{E}}, -\frac{k_p}{2} A_Z^{-1} C_p C_p^\top A_Z^{-\top} + \frac{1}{2} A_Z^\top K_q A_Z \rangle
    \\&\hspace{1cm}
    + 2\langle V_{\bar{E}}, -k_p V_{\bar{E}} A_Z^{-1} C_p C_p^\top A_Z^{-\top} \rangle, \\
    &= - \langle V_{\bar{E}}^\top V_{\bar{E}}, k_p A_Z^{-1} C_p C_p^\top A_Z^{-\top} + A_Z^\top K_q A_Z \rangle, \\
    &= - k_p \vert V_{\bar{E}} A_Z^{-1} C_p \vert^2
    - \vert A_Z V_{\bar{E}}^\top \vert_{K_q}^2.
\end{align*}
\end{proof}

Note that the statement of Lemma \ref{lem:gps_correction_V} (as well as those of Lemmas \ref{lem:gps_correction_R}-\ref{lem:mag_correction}) ignores the boundedness and uniform continuity of the correction terms to focus only on the individual contribution of each measure to the Lyapunov function decrease.
However, these conditions will be considered later on when combining the terms in Section \ref{sec:full_observer_designs}.

\begin{lemma}\label{lem:gps_correction_R}
Consider the cost function $\Lyap$ as defined in \eqref{eq:lyapunov_function}.
Choose a gain $k_c > 0$ and define the correction terms
\begin{align*}
    \Omega_\Delta &= 4 k_c R_Z^\top (\hat{y} - V_Z A_Z^{-1} C_p) \times (y - V_Z A_Z^{-1} C_p)  , \\
    W_\Delta &= k_c R_Z^\top (y - \hat{y}) C_p^\top A_Z^{-\top}, \\
    \Omega_\Gamma &= 0, \\
    W_\Gamma &= - k_c  R_Z^\top (y - V_Z A_Z^{-1} C_p)C_p^\top A_Z^{-\top}, \\
    S_\Gamma &= 0,
\end{align*}
where $y = h_p(X)$ and $\hat{y} = h_p(\hat{X})$ are the true and estimated position measurements, and $C_p^\top = (0 \; 1) \in \R^{1 \times 2}$.
Then
\begin{align*}
    \dot{\Lyap}(\Delta, \Gamma)
    &\leq -2 k_c \left(\vert (R_{\bar{E}}^2 - I_3) R_Z^\top (y - V_Z A_Z^{-1} C_p) \vert
    - \vert V_{\bar{E}} A_Z^{-1} C_p \vert\right)^2.
\end{align*}
\end{lemma}

\begin{proof}
Following the same steps as in the proof of Lemma \ref{lem:gps_correction_V}, one has that
\begin{align*}
    \dot{V}_{\bar{E}}
    &= - k_c V_{\bar{E}} A_Z^{-1} C_p C_p^\top A_Z^{-\top}.
\end{align*}
Observe that
\begin{align*}
    R_Z^\top (y - V_Z A_Z^{-1} C_p)
    &= R_Z^\top (V C_p - V_Z A_Z^{-1} C_p), \\
    &= V_{\bar{E}} A_Z^{-1} C_p
    + R_{\bar{E}} R_Z^\top (\hat{V} A_Z - V_Z) A_Z^{-1} C_p, \\
    &= V_{\bar{E}} A_Z^{-1} C_p
    + R_{\bar{E}} R_Z^\top (\hat{y} - V_Z A_Z^{-1} C_p),
\end{align*}
by the definition of $V_{\bar{E}}$ \eqref{eq:expanded_error_V}.
Thus, applying Lemma \ref{lem:rotation_error_identity}, one has that
\begin{align*}
    \dot{\Lyap}(\Delta, \Gamma)
    &= \tr(R_{\bar{E}} \Omega_\Delta^\times)
    + 2\langle V_{\bar{E}}, (I - R_{\bar{E}})W_\Gamma - R_{\bar{E}} W_\Delta \rangle, \\
    &= - 2 k_c \vert (I_3 - R^2) R_Z^\top (y - V_Z A_Z^{-1} C_p) \vert^2
    - 2 k_c \vert V_{\bar{E}} A_Z^{-1} C_p \vert^2
    \\ &\hspace{1cm}
    + 4 k_c \langle (I_3 - R^2) R_Z^\top (y - V_Z A_Z^{-1} C_p), V_{\bar{E}} A_Z^{-1} C_p \rangle \\
    &\leq - 2k_c \left(\vert (R_{\bar{E}}^2 - I_3) R_Z^\top (y - V_Z A_Z^{-1} C_p) \vert - \vert V_{\bar{E}} A_Z^{-1} C_p \vert\right)^2,
\end{align*}
as required.
\end{proof}

\subsubsection{Correction Terms using GNSS Velocity}

\begin{lemma} \label{lem:velocity_correction}
Consider the cost function $\Lyap$ as defined in \eqref{eq:lyapunov_function}.
Choose gains $k_v, k_d \geq 0$ and define the correction terms
\begin{align*}
    \Omega_\Delta &= 4 k_d R_Z^\top (\hat{y}_v - V_Z A_Z^{-1} C_v) \times (y_v - V_Z A_Z^{-1} C_v)  , \\
    W_\Delta &= (k_v+k_d) R_Z^\top (y_v - \hat{y}_v) C_v^\top A_Z^{-\top}, \\
    \Omega_\Gamma &= 0, \\
    W_\Gamma &= - (k_v+k_d)  R_Z^\top (y_v - V_Z A_Z^{-1} C_v)C_v^\top A_Z^{-\top}, \\
    S_\Gamma &= -\frac{k_v}{2} A_Z^{-1} C_v C_v^\top A_Z^{-\top},
\end{align*}
where $y_v = h_v(X)$ and $\hat{y}_v = h_v(\hat{X})$ are the true and estimated velocity measurements, and $C_v^\top = (1 \; 0) \in \R^{1 \times 2}$.
Then
\begin{align*}
    &\dot{\Lyap}(\Delta, \Gamma)
    \leq - k_v \vert V_{\bar{E}} A_Z^{-1} C_v \vert^2
    - 2 k_d \left(\vert (R_{\bar{E}}^2 - I_3) R_Z^\top (y_v - V_Z A_Z^{-1} C_v) \vert - \vert V_{\bar{E}} A_Z^{-1} C_v \vert\right)^2.
\end{align*}
\end{lemma}

\begin{proof}
This lemma is straightforward to prove by following the same steps as in the proofs of Lemmas \ref{lem:gps_correction_V} and \ref{lem:gps_correction_R}.
\end{proof}

\subsubsection{Correction Terms using Magnetometer}

\begin{lemma}\label{lem:mag_correction}
Consider the cost function $\Lyap$ as defined in \eqref{eq:lyapunov_function}.
Choose a gain $k_m > 0$ and define the correction terms
\begin{gather*}
    \Omega_\Delta = 4 k_m R_Z^\top (\hat{R} y_m)^\times \mr{y}_m  , \\
    W_\Delta = 0, \;
    \Omega_\Gamma = 0, \;
    W_\Gamma = 0, \;
    S_\Gamma = 0,
\end{gather*}
where $y_m = h_m(X)$ and $\mr{y}_m = h_m(I)$ are the measured and reference magnetic field directions \eqref{eq:magnetometer_measurement}.
Then,
\begin{align*}
    \dot{\Lyap}(\Delta, \Gamma)
    &= -2 k_m \vert (R_{\bar{E}}^2 - I_3) R_Z^\top\mr{y}_m\vert^2.
\end{align*}
\end{lemma}

\begin{proof}
Observe that
\begin{align*}
    R_Z^\top \mr{y}_m
    = R_Z^\top R R^\top \mr{y}_m
    = R_Z^\top R \hat{R}^\top R_Z R_Z^\top \hat{R} y_m
    = R_{\bar{E}} R_Z^\top \hat{R} y_m,
\end{align*}
by the definition of $R_{\bar{E}}$ \eqref{eq:expanded_error_R}.
Thus, applying Lemma \ref{lem:rotation_error_identity}, one has that
\begin{align*}
    \dot{\Lyap}(\Delta, \Gamma)
    &= \ddt \tr(I_3 - R_{\bar{E}}) + \vert V_{\bar{E}} \vert^2, \\
    &= \ddt \tr(I_3 - R_{\bar{E}})
    = -2k_m \vert (I_3 - R_{\bar{E}}^2) R_Z^\top \mr{y}_m \vert^2,
\end{align*}
as required.
\end{proof}

\section{Observer Analysis}
\label{sec:full_observer_designs}

The following theorem combines the correction terms discussed so far into one unified INS observer.
The proposed observer could be implemented without relying on GNSS velocity or on magnetometer measurements by setting the corresponding gains to zero.
Thanks to Theorem \ref{thm:combined_corrections}, the Lyapunov function decrease is guaranteed by using any combination of the individual correction terms, and the combined observer inherits the convergence and stability properties from its component parts.
It is shown that only the GNSS position is necessary for convergence and stability.

\begin{theorem}\label{thm:full_observer_design}
Let $X \in \SE_2(3)$ denote the system state with dynamics given by \eqref{eq:system_dynamics_lg}, and let $\hat{X} \in \SE_2(3)$ and $Z \in \SIM_2(3)$ denote the state estimate and auxiliary state, respectively, with dynamics given by \eqref{eq:observer_architecture}.
Let $\bar{E}$ denote the observer error as in \eqref{eq:error_definition}.
Choose gains $k_p, k_c > 0, K_q \in \Sym_+(2)$ and $k_v, k_d, k_m \geq 0$, and define the correction terms
\begin{subequations}\label{eq:full_observer_cor}
\begin{align}
    \Omega_\Delta &= 4 k_c R_Z^\top (\hat{y}_p - V_Z A_Z^{-1} C_p)^\times (y_p - V_Z A_Z^{-1} C_p)
    \notag\\ &\hspace{0.2cm} + 4 k_d R_Z^\top (\hat{y}_v - V_Z A_Z^{-1} C_v)^\times (y_v - V_Z A_Z^{-1} C_v)
    \notag\\ &\hspace{0.2cm} + 4 k_m R_Z^\top (\hat{R} y_m)^\times \mr{y}_m
    , \\
    W_\Delta &= (k_p + k_c) R_Z^\top (y_p - \hat{y}_p) C_p^\top A_Z^{-\top}
    \notag\\ &\hspace{0.2cm} + (k_v+k_d) R_Z^\top (y_v - \hat{y}_v) C_v^\top A_Z^{-\top}
    , \\
    \Omega_\Gamma &= 0, \\
    W_\Gamma &= - (k_p + k_c)  R_Z^\top (y_p - V_Z A_Z^{-1} C_p)C_p^\top A_Z^{-\top}
    \notag\\ &\hspace{0.2cm} - (k_v+k_d)  R_Z^\top (y_v - V_Z A_Z^{-1} C_v)C_v^\top A_Z^{-\top}
    , \\
    S_\Gamma &=
    \frac{1}{2} A_Z^\top K_q A_Z
    -\frac{k_p}{2} A_Z^{-1} C_p C_p^\top A_Z^{-\top}
    \notag\\ &\hspace{0.2cm}
    -\frac{k_v}{2} A_Z^{-1} C_v C_v^\top A_Z^{-\top},
\end{align}
\end{subequations}
where $y_p = h_p(X)$ and $\hat{y}_p = h_p(\hat{X})$ are the true and estimated position measurements \eqref{eq:gnss_measurement_pos}, $y_v = h_v(X)$ and $\hat{y}_v = h_v(\hat{X})$ are the true and estimated position measurements \eqref{eq:gnss_measurement_vel}, $y_m = h_m(X)$ is the true magnetometer measurement \eqref{eq:magnetometer_measurement}, and $C_p = (0 \; 1)^\top, C_v = (1 \; 0)^\top \in \R^{2 \times 1}$.
Suppose that the input $U$ is bounded, the state $X$ is uniformly continuous and bounded, and the vectors
\begin{align*}
    \mu_p &:= \sqrt{k_c} R_Z^\top (y_p - V_Z A_Z^{-1} C_p), \\
    \mu_v &:= \sqrt{k_d} R_Z^\top (y_v - V_Z A_Z^{-1} C_v), \\
    \mu_m &= \sqrt{k_m} R_Z^\top \mr{y}_m
\end{align*}
are directionally persistently exciting according to \eqref{eq:persistently_exciting}.
Then,
\begin{enumerate}
    \item \label{itm:Zboundedness} The correction terms \eqref{eq:full_observer_cor} are bounded and uniformly continuous in time.
    \item \label{itm:velocity_stability} \changeVersionOne{The origin of the combined velocity and position error $V_{\bar{E}}$ is globally exponentially stable}.
    \item \label{itm:attitude_stability} The attitude error $R_{\bar{E}}$ is almost-globally asymptotically and locally exponentially stable to the identity.
    The stable and unstable equilibria of $\bar{E}$ are given by
    \begin{align*}
        \calE_s &= \{ I_5 \}, &
        \calE_u &= \set{ \bar{E} \in \SE_2(3) }{\tr(R_{\bar{E}}) = -1}.
    \end{align*}
    \item \label{itm:estimate_convergence} If the error converges to the identity, $\vert \bar{E} - I_5 \vert \to 0$, then the state estimate converges to the system state, $\vert X - \hat{X} \vert \to 0$.
\end{enumerate}
\end{theorem}

\begin{proof}
Observe that the proposed correction terms are simply the sum of the correction terms proposed in Lemmas \ref{lem:gps_correction_V}, \ref{lem:gps_correction_R}, \ref{lem:velocity_correction}, and \ref{lem:mag_correction}.
It follows from Theorem \ref{thm:combined_corrections} that the cost function \eqref{eq:lyapunov_function} is a Lyapunov function for the error system, and its equilibrium set is the intersection of the individual equilibrium sets.
Note that this Theorem requires only that $k_p, k_c > 0, K_q \in \Sym_+(2)$ while the remaining gains $k_v, k_d, k_m$ may be set equal to zero; hence, this proof remains valid even when $k_v = k_d = k_m = 0$.

\textbf{Proof of item \ref{itm:Zboundedness}:}
To show that the correction terms \eqref{eq:full_observer_cor} are bounded and uniformly continuous in time, it suffices to show that $Z$ is uniformly continuous and well-conditioned.
We first show that $\Vert A_Z \Vert$ and $\Vert A_Z^{-1} \Vert$ are lower bounded.
To this end, let $P = A_Z A_Z^\top$, \changeVersionOne{and recall the dynamics of the components of $Z$ from Lemma \ref{lem:auxiliary_state_dynamics}.}
Then,
\begin{align}
    \label{eq:riccati_dynamics}
    \dot{P}
    &= \dot{A}_Z A_Z^\top + A_Z \dot{A}_Z^\top, \notag \\
    &= (S_N A_Z - A_Z S_\Gamma) A_Z^\top + A_Z (S_N A_Z - A_Z S_\Gamma)^\top, \notag \\
    &= S_N A_Z A_Z^\top - A_Z (S_\Gamma+S_\Gamma^\top) A_Z^\top + A_Z A_Z^\top S_N^\top, \notag \\
    &= S_N P + P S_N^\top + k_p C_p C_p^\top + k_v C_v C_v^\top - P K_q P, \notag \\
    &= S_N P + P S_N^\top + \diag(k_v, k_p) - P K_q P.
\end{align}
This is a continuous differential Riccati equation associated with state dynamics and measurement matrices,
\begin{align*}
    \left(-S_N^\top, \begin{pmatrix}
        k_p C_p^\top \\ k_v C_v^\top
    \end{pmatrix}\right)
    &= \left(
        \begin{pmatrix}
            0 & 0 \\ 1 & 0
        \end{pmatrix}, \;
        \begin{pmatrix}
            0 & k_p \\
            k_v & 0
        \end{pmatrix}
    \right).
\end{align*}
This pair is easily verified to be observable, even if $k_v = 0$, and it follows that the eigenvalues of $ P$ and hence of $ A_Z$ are bounded above and below by some nonzero constants.

Since $R_Z \in \SO(3)$ is constant and bounded by definition and $A_Z$ was shown to be well-conditioned, it remains only to show that $V_Z$ is bounded above.
To see this is so, recall \eqref{eq:expanded_auxiliary_dynamics_V} and let $l = \vert V_Z \vert^2$.
Then,
\begin{align*}
    &\dot{l}
    = 2 \langle V_Z, W_G A_Z - R_Z W_\Gamma - V_Z S_\Gamma \rangle, \\
    &= \langle V_Z, 2 W_G A_Z + (k_p + k_c) y_p C_p^\top A_Z^{-\top}
    + (k_v+k_d) y_v C_v^\top A_Z^{-\top} \rangle
    \\ &\hspace{1cm}
    - \langle V_Z V_Z^\top, A_Z^\top K_q A_Z + (k_p + k_c) A_Z^{-1} C_p C_p^\top A_Z^{-\top}
    + (k_v+k_d) A_Z^{-1} C_p C_p^\top A_Z^{-\top}
     \rangle, \\
    &\leq \vert 2 W_G A_Z + (k_p + k_c) y_p C_p^\top A_Z^{-\top}
    + (k_v+k_d) y_v C_v^\top A_Z^{-\top} \vert \sqrt{l}
    \\ &\hspace{1cm}
    - 2\lambda_{\min}(A_Z^\top K_q A_Z) l.
\end{align*}
Since the coefficient of $\sqrt{l}$ is bounded, it follows that $l$ and hence $V_Z$ is bounded above.

\textbf{Proof of item \ref{itm:velocity_stability}:}

Let $\underline{\alpha} > 0$ denote the lower bound of  $\Vert A_Z \Vert$ and let $\underline{q}$ denote the least eigenvalue of $K_q$.
Then, by Lemma \ref{lem:gps_correction_V}, the combined velocity and position error $V_{\bar{E}}$ satisfies
\begin{align*}
    \ddt \vert V_{\bar{E}} \vert^2
    \leq - 2\vert A_Z V_{\bar{E}}^\top \vert^2_{K_q}
    \leq - 2\underline{\alpha}^2 \underline{q} \vert V_{\bar{E}} \vert^2.
\end{align*}
This shows that $V_{\bar{E}}$ is globally exponentially stable to zero.

\textbf{Proof of item \ref{itm:attitude_stability}:}
We apply Barbalat's lemma to verify that $R_{\bar{E}} \to I_3$.
From item \ref{itm:velocity_stability}, it holds that $V_{\bar{E}} \to 0$ exponentially.
Combining this with the equilibrium sets obtained from Lemmas \ref{lem:gps_correction_R}, \ref{lem:velocity_correction}, \ref{lem:mag_correction}, one has that
\begin{align*}
    \dot{\Lyap} &\to - 2k_c \vert (R_{\bar{E}}^2 - I_3) R_Z^\top (y - V_Z A_Z^{-1} C_p) \vert^2
    \\ &\hspace{1cm}
    - 2k_d \vert (R_{\bar{E}}^2 - I_3) R_Z^\top (y - V_Z A_Z^{-1} C_v) \vert^2
    \\ &\hspace{1cm}
    - 2k_m \vert (R_{\bar{E}}^2 - I_3) R_Z^\top \mr{y}_m \vert^2
    \to 0,
\end{align*}
since $\dot{\Lyap}$ is uniformly continuous as the sum and product of uniformly continuous and bounded signals.
Thus,
\begin{align*}
    R_{\bar{E}}^2 \mu_p &\to \mu_p, &
    R_{\bar{E}}^2 \mu_v &\to \mu_v, &
    R_{\bar{E}}^2 \mu_m &\to \mu_m.
\end{align*}
By assumption, the combination of these vectors is directionally persistently exciting and therefore $R_{\bar{E}} \to I_3$ or $R_{\bar{E}} \to U \Lambda U^\top$ where $U \in \SO(3)$ and $\Lambda = \diag(1,-1,-1)$ \cite[Theorem 4.3]{2012_trumpf_AnalysisNonLinearAttitude}.

In the case that $\bar{E} \to (I_3, 0_{3\times 2}) \in \calE_s$, we show that the desired equilibrium is locally exponentially stable.
By linearising the dynamics of $R_{\bar{E}}$ about $R_{\bar{E}} \approx I_3 + \bar{\varepsilon}_R^\times$ and $V_{\bar{E}} \approx 0$, one has
\begin{align*}
    \dot{\bar{\varepsilon}}_{R}
    &\approx - 4 \sum_{s \in \{p,v,m\}} \mu_s \times ((I_3 + \bar{\varepsilon}_R^\times)\mu_s), \\
    &= - 4 \sum_{s \in \{p,v,m\}} \mu_s^\times \bar{\varepsilon}_R^\times \mu_s
    = 4 \sum_{s \in \{p,v,m\}} \mu_s^\times \mu_s^\times \bar{\varepsilon}_R.
\end{align*}
Since $\sum_{s \in \{p,v,m\}} \mu_s^\times \mu_s^\times$ is negative semi-definite and persistently exciting, this confirms that $\bar{\varepsilon}_R$ is (uniformly) locally exponentially stable by \cite[Theorem 1]{1977_morgan_UniformAsymptoticStability}.

In the case that $\bar{E} \to (U \Lambda U^\top, 0_{3\times 2}) \in \calE_u$, we show that this equilibrium is unstable by identifying an infinitesimally close point in $\SE_2(3)$ for which the value of the Lyapunov function is smaller.
Consider a fixed but arbitrary equilibrium $\bar{E} = (U \Lambda U^\top, 0) \in \calE_u$ of $\dot{\Lyap}(\Delta, \Gamma)$ and define $Q(s) = U \Lambda U^\top \exp(s (U \eb_1)^\times)$.
Then, one applies a second-order Taylor expansion for $s$ to find that
\begin{align*}
    \Lyap(Q(s), 0)
    &= \tr(I_3 - U \Lambda U^\top \exp(s (U \eb_1)^\times)), \\
    &\approx \tr(I_3 - U \Lambda U^\top (I_3 + s (U \eb_1)^\times
    + \frac{s^2}{2} (U \eb_1)^\times(U \eb_1)^\times)), \\
    &= 4 - s \tr(U \Lambda U^\top ( U \eb_1)^\times)
    - \frac{s^2}{2} \tr(U \Lambda U^\top ( U \eb_1)^\times(U \eb_1)^\times), \\
    &= 4 - \frac{s^2}{2} \tr(U \Lambda U^\top ( U \eb_1)^\times(U \eb_1)^\times)
    = 4 - \frac{s^2}{2} \tr(\Lambda \eb_1^\times \eb_1^\times), \\
    &= 4 - \frac{s^2}{2} (\eb_1^\top \Lambda \eb_1  - \tr(\Lambda) )
    = 4 - s^2.
\end{align*}
Thus there exists $(Q(s), 0)$ for which $\Lyap(Q(s),0) < \Lyap(U \Lambda U^\top, 0) = \Lyap(\bar{E})$ in any neighbourhood of the equilibrium $\bar{E} \in \calE_u$.
Since $\bar{E} \in \calE_u$ was chosen arbitrarily, it follows that $\calE_u$ is the set of unstable equilibria.

\changeVersionTwo{
To show the almost global nature of the asymptotic stability of $R_{\bar{E}} \to I_3$, assume without loss of generality that $V_{\bar{E}}\equiv 0$.
Recall that the set of unstable points is characterized by $R_{\bar{E}} =U \Lambda U^\top$ or $\tr(R_{\bar{E}})=-1$.
Its complement, defined by $\tr(R_{\bar{E}})\neq-1$, is open and dense in $\SO(3)$.
Therefore, for all initial conditions $R_{\bar{E}} \in \SO(3)\setminus\{\tr(R_{\bar{E}})=-1\}$, $R_{\bar{E}}\to I_3 $.
Note that the set of unstable equilibria is the set of all $180$-degree rotations, each defined by a unique axis of rotation.
These axes form points on $\Sph^2$, the 2-dimensional unit sphere.
Hence, the set of unstable points forms a 2-dimensional manifold embedded in the 3-dimensional space of rotations $\SO(3)$.
This shows that the set of the unstable set has measure zero in $\SO(3)$.
}

\textbf{Proof of item \ref{itm:estimate_convergence}:}

The proof is completed by applying \cite[Lemma 5.3]{2021_vangoor_AutonomousErrorConstructive}.
Since $\bar{E}:= Z^{-1} X \hat{X}^{-1} Z \to I_5$ and $Z$ is well conditioned, one directly concludes that $X \hat{X}^{-1}\to I_5$, which implies that $X -\hat{X} \to 0$.
\end{proof}

\begin{remark}
The requirement of persistence of excitation in Theorem \ref{thm:full_observer_design} is generally easy to satisfy when the magnetometer is used.
A set of two time-varying vectors is directionally persistently exciting as long as they are uniformly non-colinear.
The vectors $\mu_p$ and $\mu_v$ may become aligned when the vehicle travels in a straight line for a long time.
However, it is highly unusual that the magnetometer vector $\mu_m$ is also aligned with $\mu_p$ and $\mu_v$.
These observations are reflected in the simulation results.
\end{remark}

\begin{remark}
The observer gains and the initial conditions of $A_Z$ can be chosen such that $\dot{A}_Z = 0$ for all time.
Doing so yields a simplified observer where only $V_Z A_Z^{-1} \in \R^{3\times 2}$ needs to be tracked, rather than tracking $A_Z$ and $V_Z$ separately.
This simplified observer is closely linked to the design proposed in the authors' prior work \cite{2023_vangoor_ConstructiveEquivariantObserverb}, although it also includes the GNSS velocity and magnetometer corrections while \cite{2023_vangoor_ConstructiveEquivariantObserverb} did not.
\end{remark}

\changeVersionOne{
\subsection{Selection of Observer Gains}

The proposed observer has a number of gains that can be tuned to adjust its performance.
The gains $k_p, k_c, k_v, k_d, k_m$ are each associated with particular sensors, as described in Table \ref{tab:gain_explanation_table}.
Increasing any of these gains generally increases the rate of convergence but also the observer's sensitivity to the associated sensors.
The gain $K_q$ is not associated with a particular sensor but is related to the `relaxation' of the position and velocity estimates.
Increasing $K_q$ makes the position and velocity estimate more sensitive to measurements while increasing only the top-left component of $K_q$ makes the velocity estimate less sensitive to position measurements.
The scaling component $A_Z$ of the auxiliary state behaves like a time-varying gain inside the observer, similar to a Riccati equation (see \eqref{eq:riccati_dynamics}).
A large initial value for $A_Z$ increases the observer's initial sensitivity and can reduce the time it takes to reach the steady-state response.
}

\begin{table}[tb]
    \changeVersionOne{
    \caption{Descriptions of the gains of the proposed observer.}
    \centering
        \begin{tabularx}{0.5\linewidth}{|x{\widthof{Gains}}|x{\widthof{Magnetometer}}|X|}
            \hline
            Gain & Sensor & State Estimates Most Affected \\
            \hline
            $k_p$ & GNSS Pos. &
            Position and velocity \\
            $k_c$ & GNSS Pos. &
            Attitude (mostly pitch and roll) \\
            $k_v$ & GNSS Vel. &
            Velocity \\
            $k_d$ & GNSS Vel. &
            Attitude (mostly pitch and roll) \\
            $k_m$ & Magnetometer &
            Attitude (mostly yaw) \\
            \hline
        \end{tabularx}
    \label{tab:gain_explanation_table}
    }
\end{table}

\section{Simulations}

Simulations of a flying vehicle were used to verify the proposed observer design in Theorem \ref{thm:full_observer_design}.
The vehicle was simulated as flying in a circle of radius 50~m with a constant velocity of 25~m/s.
The initial attitude, velocity, and position were set to
\begin{align*}
    R(0) &= I_3, &
    v(0) &= 25.0\eb_2\text{~m/s}, &
    p(0) &= 50.0\eb_1\text{~m}.
\end{align*}
The accelerometer and gyroscope inputs were defined as
\begin{align*}
    \Omega &= \eb_3\text{~rad/s}, &
    a &= -0.25 R^\top p - R^\top g \text{~m/s$^2$},
\end{align*}
with the inertial gravity vector defined by $g = 9.81\eb_3$~m/s$^2$.
The reference magnetic field direction was defined to be $\mr{y}_m = \eb_1\text{~$\mu$T}$.

The observer proposed in Theorem \ref{thm:full_observer_design} was implemented in four versions, with each using a different combination of sensors.
The observers are labeled with a substring of `pvm' depending on the sensors used; for example, the observer using \textbf{p}osition and \textbf{m}agnetometer measurements is labeled `pm', and this is implemented by setting $k_v=k_d=0$.
The gains for each variation of the observer are shown in Table \ref{tab:gain_table}.

\begin{table}[!htb]
    \centering
    \caption{The gains used for the four observers in simulation.}
    \begin{tabular}{|c||c|c|c|c|c|c|}
        \hline
        Observer & $K_q$      & $k_p$ & $k_c$ & $k_v$ & $k_d$ & $k_m$ \\
        \hline
        Est. pvm & $\diag(10.0,2.0)$ & 10.0 & 0.1 & 10.0 & 0.1 & 2.0 \\
        Est. pv  & $\diag(10.0,2.0)$ & 10.0 & 0.1 & 10.0 & 0.1 & 0.0 \\
        Est. pm  & $\diag(10.0,2.0)$ & 10.0 & 0.1 & 0.0 & 0.0 & 2.0  \\
        Est. p   & $\diag(10.0,2.0)$ & 10.0 & 0.1 & 0.0 & 0.0 & 0.0  \\
        \hline
    \end{tabular}
    \label{tab:gain_table}
\end{table}

The four observer versions were provided with an extreme initial condition to demonstrate their almost-global stability.
Specifically, the initial condition of each observer version was defined by
\begin{align*}
    \hat{R}(0) &= \exp(0.99\pi \eb_1^\times), &
    \hat{v}(0) &= \begin{pmatrix}
        2 & 27 & 2
    \end{pmatrix}^\top\text{~m/s}, &
    \hat{p}(0) &= \begin{pmatrix}
        70 & 20 &20
    \end{pmatrix}^\top\text{~m}.
\end{align*}
The auxiliary states were also initialised with the same value for each observer version,
\begin{align*}
    R_Z(0) &= I_3, &
    A_Z(0) &= \diag(2, 10), &
    V_Z(0) &= \hat{V}(0) A_Z(0).
\end{align*}
The system and observer dynamics were simulated using Lie group Euler integration at 50~Hz for 50~s.

\begin{figure*}
\includegraphics[width=1.0\linewidth]{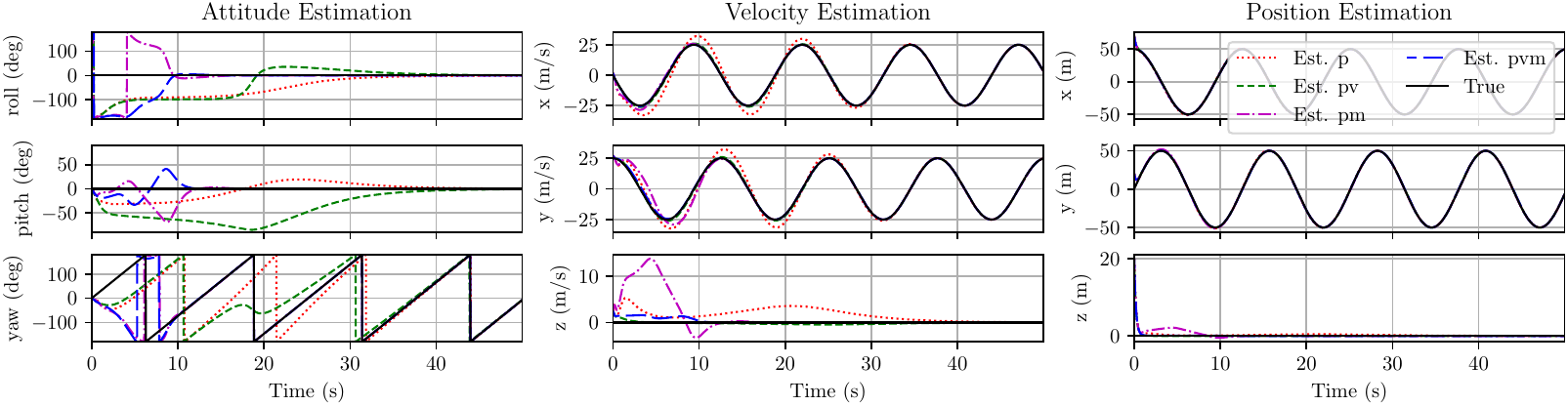}
\caption{The estimated navigation states from the proposed observer using the four different sets of gains
(p:\textcolor{red}{$\cdots$}, pv:\textcolor{green}{- - -}, pm:\textcolor{purple}{-$\cdot$-$\cdot$}, pvm:\textcolor{blue}{--- ---})
described in Table \ref{tab:gain_table} are shown alongside the true (\textbf{-----}) navigations states.
}
\label{fig:INS_estimation_extreme}
\end{figure*}

\begin{figure}
    \centering
    \includegraphics[width=0.6\linewidth]{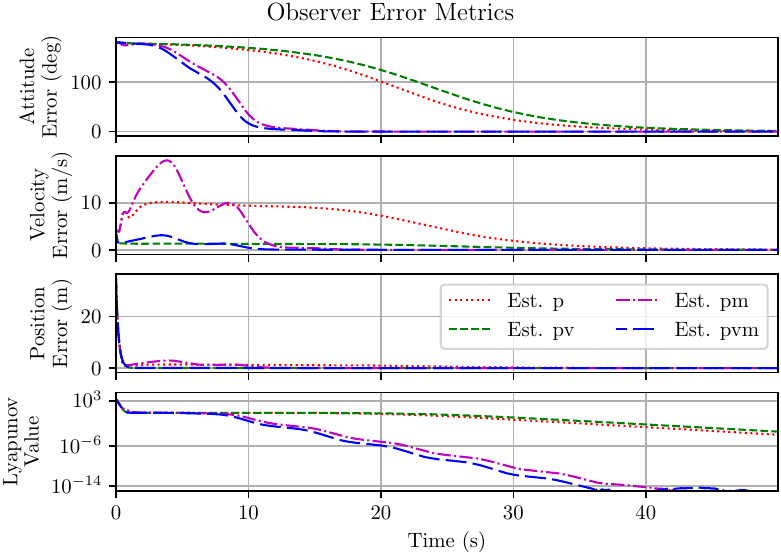}
    \caption{The errors in the estimated navigation states and the Lyapunov function value of the proposed observer using the four different sets of gains
    (p:\textcolor{red}{$\cdots$}, pv:\textcolor{green}{- - -}, pm:\textcolor{purple}{-$\cdot$-$\cdot$}, pvm:\textcolor{blue}{--- ---})
    described in Table \ref{tab:gain_table}.
    }
    \label{fig:INS_observer_error_extreme}
\end{figure}

Figure \ref{fig:INS_estimation_extreme} shows the estimated navigation states obtained from the four observer variants along with the true states of the simulated system.
Figure \ref{fig:INS_observer_error_extreme} shows the absolute errors in attitude, position, and velocity, in addition to the value of the Lyapunov function for each of the observers.
One notes that the Lyapunov function is monotonically decreasing for each observer, which verifies Theorem \ref{thm:full_observer_design}.
Additionally, each observer's estimated states converge to the true states in spite of the extreme initial condition.

The advantage in using other measurements in addition to GNSS position is apparent in the convergence rates of the various states.
Shown in the attitude estimation plots of Figure \ref{fig:INS_estimation_extreme} and the attitude error plot of Figure \ref{fig:INS_observer_error_extreme}, the yaw estimates of observers that utilise the magnetometer measurement \eqref{eq:magnetometer_measurement} are seen to converge far more quickly than the observers that do not.
Likewise, the velocity estimates of the observers that utilise the GNSS velocity measurement \eqref{eq:gnss_measurement_vel} exhibit a far smaller initial error and converge more quickly than those of the observers that do not.
It is interesting to compare the magnetometer `pm' and velocity `pv' observers; under the chosen simulation conditions, the `pm' observer converges faster overall while the `pv' observer incurs a much smaller initial error in both position and velocity.
Understanding these behaviours can be useful for practitioners who may need to make tradeoffs between hardware based on typical operation scenarios.

What is perhaps surprising is that the attitude estimate of the `pv' observer is slower to converge than the `p' observer.
This can be explained by the tradeoff between convergence of the position and velocity and damping of the persistence of excitation signal required to converge attitude when the magnetometer is not available.
We believe that the behaviour of the `pv' observer could be improved through more careful tuning of the gains,
however, the gains here were chosen to be simple and similar across the observer designs to provide a demonstration of relative observer behaviour rather than tuned for a specific scenario.

\section{Experiments}

A real-world experiment was carried out using a remotely piloted aerial system (RPAS) equipped with the ArduPilot open source flight controller \cite{2010_tridgell_ArduPilot}.
ArduPilot was used to record data from the IMU (350~Hz), magnetometer (10~Hz), and GNSS (5~Hz) over a period of six minutes while the plane flew around an area of approximately 400~m $\times$ 400~m.
The GNSS measurements were converted from Latitude-Longitude-Altitude (LLA) into a local North-East-Down (NED) frame by using the first GNSS position measurement as a reference.
The reference magnetic field was obtained from the ArduPilot Python tools as $\mr{y}_m = (23.33, 5.19, -52.80)$~$\mu$T.
ArduPilot's EKF3 is used for comparison with our proposed observer.
It is based on a multiplicative Kalman filter \cite{2017_sola_QuaternionKinematicsErrorstate} but also includes a range of initialisation and outlier rejection techniques to mitigate issues related to system nonlinearities and non-Gaussian measurement noise.
The proposed observer was implemented at the rate of the IMU measurements, and the latest magnetometer and GNSS measurements were used as if they were current, i.e. the GNSS and magnetometer measurements were treated as piecewise constant in a zero-order hold scheme.
The initial conditions were chosen as
\begin{align*}
    \hat{R}(0) &= I_3, &
    \hat{v}(0) &= 0_{3\times1}\text{~m/s}, &
    \hat{p}(0) &= 0_{3\times1}\text{~m}, \\
    \hat{R}_Z(0) &= I_3, &
    \hat{V}_Z(0) &= 0_{3\times 2}, &
    \hat{A}_Z(0) &= I_2,
\end{align*}
and the gains were set to
\begin{align*}
    K_q &= \diag(0.1, 0.02), &
    k_p &= 1.0, &
    k_c &= 0.01, \\
    k_v &= 1.0, &
    k_d &= 0.001, &
    k_m &= 2.0\times10^{-6}.
\end{align*}

\begin{figure*}
    \includegraphics[width=1.0\linewidth]{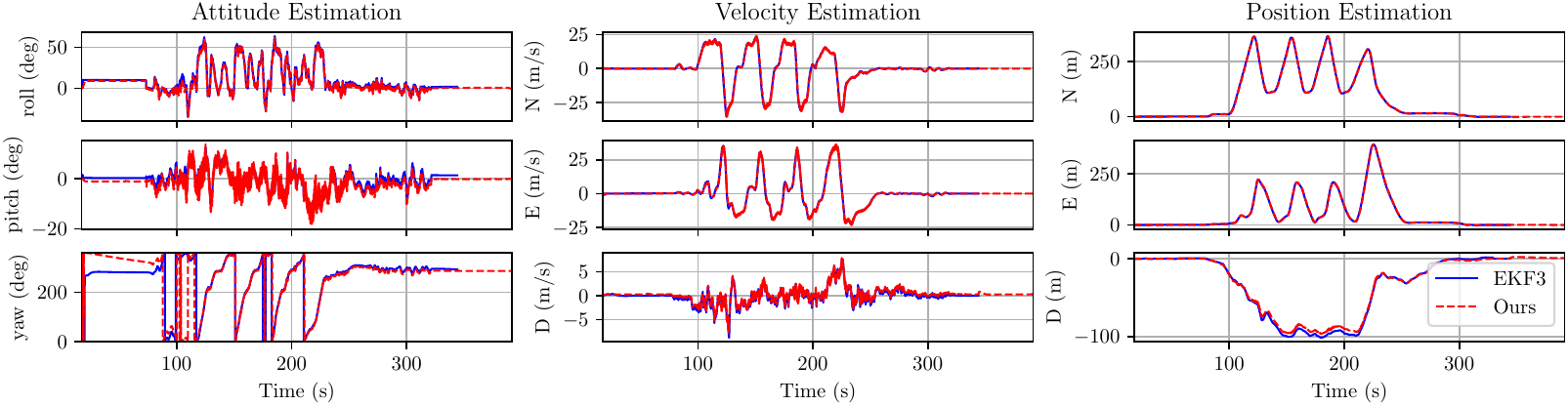}
    \caption{The estimated navigation states from the proposed observer (\textcolor{red}{- - -}) and the ArduPilot EKF3 (\textcolor{blue}{-----}) over the whole experimental flight.
    The trajectory estimates are very similar.}
    \label{fig:INS_estimation_real_full}
\end{figure*}

\begin{figure*}
    \includegraphics[width=1.0\linewidth]{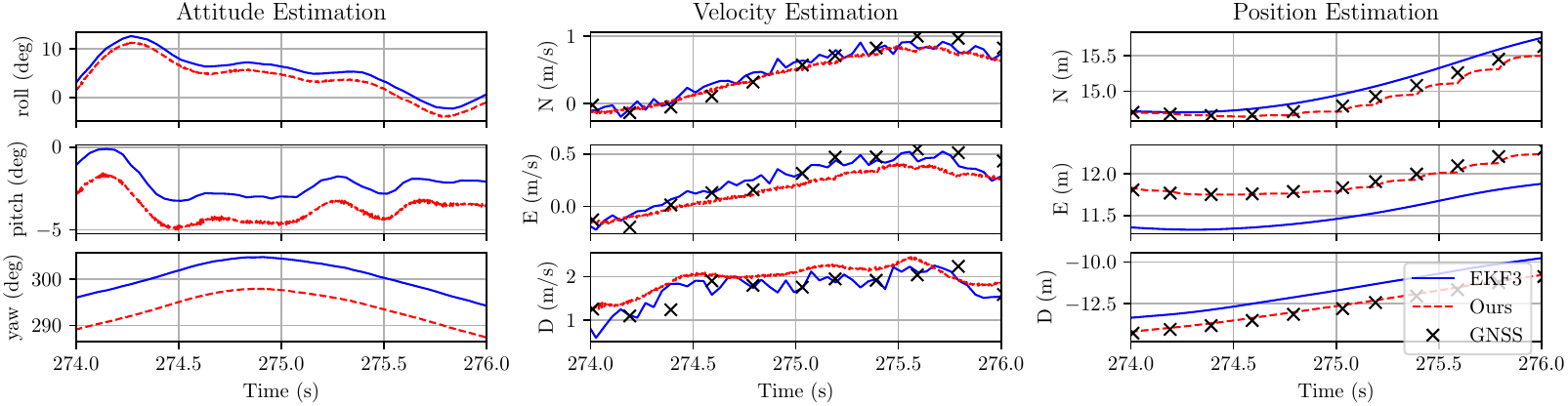}
    \caption{The estimated navigation states from the proposed observer (\textcolor{red}{- - -}) and the ArduPilot EKF3 (\textcolor{blue}{-----}) and the GNSS measurements ($\times \times \times$) over a two second period of the experimental flight.}
    \label{fig:INS_estimation_real_brief}
\end{figure*}

Figure \ref{fig:INS_estimation_real_full} shows the navigation states estimated by the proposed observer compared with those provided by the ArduPilot EKF3 state estimator.
Figure \ref{fig:INS_estimation_real_brief} additionally shows these estimates over two seconds alongside the GNSS measurements during this time.
While the plane is stationary during the first 80~s of data, the yaw estimate of the proposed observer converges to be close to that provided by the EKF3.
Throughout the flight, the estimated trajectories are very similar between the two observers, although there is a clear offset in the estimated heights (D-position).
This is due to the EKF3 preferring the barometer as a measurement of height over the GNSS, while the proposed observer only uses the GNSS.
In the enlarged view presented in Figure \ref{fig:INS_estimation_real_brief}, the effect of the zero-order hold scheme can be seen in the position estimates of the proposed observer.
The effect is less pronounced in the velocity estimates, where it is interesting to note that the estimate provided by our proposed observer is significantly smoother than that provided by the EKF3, \changeVersionTwo{despite the latter involving a much larger number of states%
\footnote{\changeVersionTwo{ArduPilot's EKF3 involves 299 states: 23 for the components of the state vector (which also includes estimates of gyroscope and accelerometer biases, North and East wind velocity, Earth-frame magnetic flux bias, magnetometer bias, and vehicle height above the terrain), and 12*23 for the components of the covariance matrix.}}
than the proposed observer%
\footnote{\changeVersionTwo{The proposed observer involves $25$ states: 15 for the components of $\hat{X}$ and 10 for the components of $Z$.}}}

The discrepancy in attitude estimation is likely due to biases in the gyroscope and magnetometer measurements that are not considered in the proposed observer design.
Compensating for such issues is left as an interesting open question for future work.
Regardless, the proposed observer provides a solution for attitude, velocity, and position that is comparable to the state-of-the-art ArduPilot EKF3.

\section{Conclusion}

This paper proposes, to the authors' knowledge, the first observer for INS with almost globally asymptotically and locally exponentially stable error dynamics.
These properties are obtained by exploiting the newly recognised automorphism group $\SIM_2(3)$ of the extended special Euclidean group $\SE_2(3)$ and its application to developing a synchronous observer architecture.
The resulting observer design is able to converge using only GNSS position measurements or GNSS position with a combination of GNSS velocity and magnetometer measurements.
A simulation is used to demonstrate the observer convergence of various gain configurations from an extreme initial condition with an angle error of $0.99\pi$~rad, and
a real-world experiment shows practical performance as compared to a state-of-the-art multiplicative EKF implementation.

This work opens a number of exciting avenues for future research. A key challenge is to incorporate sensor bias estimation into the present design, an important step for future practical implementations.
Direct inclusion of barometer measurements in the modular observer architecture is also an open problem.
We believe that the convergence of the attitude estimation could be improved by exploring time-varying or non-scalar gains.
Tightly coupled GNSS measurements have been shown to improve INS performance \cite{2019_hansen_NonlinearObserverTightly} and are also an interesting direction to consider.
Overall, we hope that this paper's design principles enable further INS solutions with almost-global stability characteristics.

\section*{Acknowledgements}
This research was supported by the Australian Research Council through the Discovery Project DP210102607, the French National Research Agency through the Project ASCAR-ANR-23-ASTR-0016, the Franco-Australian International Research Project “Advancing Autonomy for Unmanned Robotic Systems” (IRP ARS), and the Horizon Europe MSCA PF MEW (101154194).

The authors would like to thank Andrew Tridgell for his invaluable support in collecting and processing experimental data from ArduPilot.

\bibliographystyle{abbrv}
\bibliography{Automatica_2023_INS}

\appendix

\begin{longversion}
\section{Lemmas and Proofs}
\end{longversion}
\begin{shortversion}
\section{Appendix}
\end{shortversion}

\begin{lemma}\label{lem:rotation_error_identity}
Suppose that $R \in \SO(3)$ and $\dot{R} = - 2 k R (\hat{\mu} \times \mu)^\times$.
If $\mu = R \hat{\mu} + \delta$, then
\begin{align*}
    \ddt \tr(I_3 - R) = -  \vert (I_3 - R^2) \mu \vert^2
    + 2 k \langle (I_3 - R^2) \mu, \delta \rangle.
\end{align*}
\end{lemma}

\begin{proof}
Direct computation yields
\begin{align*}
    \ddt \tr(I_3 - R)
    &= - \tr (\dot{R}), \\
    &= 2 \tr ( k R (\hat{\mu} \times \mu)^\times), \\
    &= -2 k \tr (R (\hat{\mu} \mu^\top - \mu \hat{\mu}^\top)), \\
    &= -2k \tr (R \hat{\mu} \mu^\top - R \mu \hat{\mu}^\top), \\
    &= -2k \tr ((\mu - \delta) \mu^\top - R \mu (R^\top \mu - R^\top \delta)^\top), \\
    &= -2k \tr (\mu \mu^\top -R \mu \mu^\top R - \delta \mu^\top + R \mu \delta^\top R), \\
    &= -2k \mu^\top(I_3 - R^2) \mu + 2 k \mu^\top (I_3 - R^2)^\top \delta, \\
    &= -k \mu^\top (I_3 - R^2)^\top (I_3 - R^2) \mu
    \\ &\hspace{1cm}
    + 2k \mu^\top (I_3 - R^2)^\top \delta, \\
    &= -k \vert (I_3 - R^2) \mu \vert^2
    + 2k \mu^\top (I_3 - R^2)^\top \delta.
\end{align*}
This completes the proof.
\end{proof}

\begin{longversion}

\begin{proof}[Proof of Lemma \ref{lem:auxiliary_state_dynamics}]
Direct computation in matrix form yields
\begin{align*}
    \dot{Z}
    &= (G + N) Z - Z \Gamma, \\
    &= \begin{pmatrix}
        0_{3\times3} & W_G \\ 0_{2\times3} & S_N
    \end{pmatrix}
    \begin{pmatrix}
        R_Z & V_Z \\ 0_{2\times3} & A_Z
    \end{pmatrix}
    - \begin{pmatrix}
        R_Z & V_Z \\ 0_{2\times3} & A_Z
    \end{pmatrix}
    \begin{pmatrix}
        \Omega_\Gamma^\times & W_\Gamma \\ 0_{2\times3} & S_\Gamma
    \end{pmatrix}, \\
    &= \begin{pmatrix}
        0_{3\times3} & W_G A_Z \\ 0_{2\times3} & S_N A_Z
    \end{pmatrix}
    - \begin{pmatrix}
        R_Z \Omega_\Gamma^\times & R_Z W_\Gamma + V_Z S_\Gamma \\ 0_{2\times3} & A_Z S_\Gamma
    \end{pmatrix}, \\
    &= \begin{pmatrix}
        -R_Z \Omega_\Gamma^\times &&&
        W_G A_Z - R_Z W_\Gamma - V_Z S_\Gamma \\
        0_{2\times3} &&&
        S_N A_Z - A_Z S_\Gamma
    \end{pmatrix}.
\end{align*}

\end{proof}

\begin{proof}[Proof of Lemma \ref{lem:error_state_and_dynamics}]
Expansion of \eqref{eq:error_definition} in matrix form yields
\begin{align}
    \bar{E}
    &= \begin{pmatrix}
        R_Z & V_Z \\ 0_{2\times 3} & A_Z
    \end{pmatrix}^{-1}
    \begin{pmatrix}
        R & V \\ 0_{2\times 3} & I_2
    \end{pmatrix}
    \begin{pmatrix}
        \hat{R} & \hat{V} \\ 0_{2\times 3} & I_2
    \end{pmatrix}
    \begin{pmatrix}
        R_Z & V_Z \\ 0_{2\times 3} & A_Z
    \end{pmatrix}, \notag \\
    &= \begin{pmatrix}
        R_Z^\top & -R_Z^\top V_Z A_Z^{-1} \\ 0_{2\times 3} & A_Z^{-1}
    \end{pmatrix}
    \begin{pmatrix}
        R & V \\ 0_{2\times 3} & I_2
    \end{pmatrix}
    \begin{pmatrix}
        \hat{R}^\top & -\hat{R}^\top \hat{V} \\ 0_{2\times 3} & I_2
    \end{pmatrix}
    \begin{pmatrix}
        R_Z & V_Z \\ 0_{2\times 3} & A_Z
    \end{pmatrix}, \notag \\
    &= \begin{pmatrix}
        R_Z^\top R & -R_Z^\top V_Z A_Z^{-1} + R_Z^\top V \\ 0_{2\times 3} & A_Z^{-1}
    \end{pmatrix}
    \begin{pmatrix}
        \hat{R}^\top R_Z & \hat{R}^\top V_Z -\hat{R}^\top \hat{V}A_Z \\ 0_{2\times 3} & A_Z
    \end{pmatrix}, \notag \\
    R_{\bar{E}} &= R_Z^\top R \hat{R}^\top R_Z, \\
    V_{\bar{E}} &= R_Z^\top R (\hat{R}^\top V_Z -\hat{R}^\top \hat{V}A_Z ) + (R_Z^\top V -R_Z^\top V_Z A_Z^{-1}) A_Z, \notag \\
    &= R_Z^\top(V A_Z - V_Z) - R_{\bar{E}} R_Z^\top (\hat{V} A_Z - V_Z).
\end{align}
Likewise, expansion of \eqref{eq:error_dynamics} in matrix form yields
\begin{align}
    \dot{\bar{E}}
    &= \begin{pmatrix}
        \Omega_\Gamma^\times & W_\Gamma \\ 0_{2\times 3} & S_\Gamma
    \end{pmatrix}
    \begin{pmatrix}
        R_{\bar{E}} & V_{\bar{E}} \\ 0_{2\times 3} & I_2
    \end{pmatrix}
    - \begin{pmatrix}
        R_{\bar{E}} & V_{\bar{E}} \\ 0_{2\times 3} & I_2
    \end{pmatrix}
    \begin{pmatrix}
        \Omega_\Gamma^\times & W_\Gamma \\ 0_{2\times 3} & S_\Gamma
    \end{pmatrix}
    - \begin{pmatrix}
        R_{\bar{E}} & V_{\bar{E}} \\ 0_{2\times 3} & I_2
    \end{pmatrix}
    \begin{pmatrix}
        \Omega_\Delta^\times & W_\Delta \\ 0_{2\times 3} & 0_{2\times 2}
    \end{pmatrix}, \notag \\
    &= \begin{pmatrix}
        \Omega_\Gamma^\times R_{\bar{E}} & \Omega_\Gamma^\times V_{\bar{E}} + W_\Gamma \\ 0_{2\times 3} & S_\Gamma
    \end{pmatrix}
    - \begin{pmatrix}
        R_{\bar{E}} \Omega_\Gamma^\times & R_{\bar{E}} W_\Gamma + V_{\bar{E}} S_\Gamma \\ 0_{2\times 3} & S_\Gamma
    \end{pmatrix}
    - \begin{pmatrix}
        R_{\bar{E}} \Omega_\Delta^\times & R_{\bar{E}} W_\Delta \\ 0_{2\times 3} & 0_{2\times 2}
    \end{pmatrix}, \notag \\
    \dot{R}_{\bar{E}} &= \Omega_\Gamma^\times R_{\bar{E}} - R_{\bar{E}}(\Omega_\Gamma^\times + \Omega_\Delta^\times), \\
    \dot{V}_{\bar{E}} &= \Omega_\Gamma^\times V_{\bar{E}} - V_{\bar{E}} S_\Gamma + (I - R_{\bar{E}})W_\Gamma - R_{\bar{E}} W_\Delta.
\end{align}
This completes the proof.
\end{proof}
\end{longversion}

\end{document}